\documentclass[nofootinbib,amsmath,amssymb,amsfonts,aps,prd,eqsecnum,superscriptaddress]{revtex4-1}

\pdfoutput=1
\usepackage{hyperref}
\usepackage{amsmath}
\usepackage{amssymb}
\usepackage{mathrsfs}
\usepackage{xcolor,graphicx}
\usepackage[caption=false]{subfig}
\usepackage[toc,page]{appendix}
\usepackage[section]{placeins}

\begin{document}

\title{Scaling Relations in Two-Dimensional Relativistic Hydrodynamic Turbulence}

\author{John Ryan Westernacher-Schneider}
\email{jwestern@uoguelph.ca}
\affiliation{Department of Physics \\
  University of Guelph \\
  Guelph, Ontario N1G 2W1, Canada}
\affiliation{Perimeter Institute for Theoretical Physics \\
  31 Caroline Street North \\
  Waterloo, Ontario N2L 2Y5, Canada}

\author{Luis Lehner}
\email{llehner@perimeterinstitute.ca}
\affiliation{Perimeter Institute for Theoretical Physics \\
  31 Caroline Street North \\
  Waterloo, Ontario N2L 2Y5, Canada}

\author{Yaron Oz}
\email{yaronoz@post.tau.ac.il}
\affiliation{Raymond and Beverly Sackler School of Physics and Astronomy,\\
Tel Aviv University, \\
Tel Aviv 69978, Israel}

\date{\today}

\begin{abstract}
We derive exact scaling relations for two-dimensional relativistic hydrodynamic turbulence in the inertial
range of scales.
We consider both the energy cascade towards large scales and the enstrophy cascade towards small scales.
We illustrate
these relations by numerical simulations of turbulent  weakly compressible
flows.
Intriguingly, the fluid-gravity correspondence implies
that the gravitational field 
in spacetimes with anti-de Sitter asymptotics 
should exhibit similar scaling relations.

\end{abstract}

\maketitle

\section{Introduction}
Turbulence is a ubiquitous phenomenon 
of fluid flows  which plays a
key role in many physical scenarios. At a broad level, turbulence takes place when non-linear 
interactions of a large number of degrees of freedom dominate over dissipative effects (the so-called high Reynolds number regime). Due to its intrinsic
strongly non-linear and far from equilibrium character, a thorough understanding of this phenomenon from first principles remains elusive. The goal of
understanding goes well beyond academic interests, as a deeper grasp of this phenomenon would
impact a broad range of areas including weather dynamics, astrophysical processes, aerodynamics, etc.

Making this enterprise difficult, as thoroughly discussed in e.g. \cite{reynolds1883experimental, Kolmogorov:1941,Kolmogorov:1941b,Batchelor:1956, Kraichnan:1967, Frisch:1996}, is turbulence's chaotic nature, the flow of energy towards
smaller or larger wavelengths, and its non-linearity.
Common approaches in the analytical study of 
fluid turbulence rely on dimensional and statistical arguments, often assuming as many statistical symmetries as are possible. These include rotational, parity, and translational invariance, as well as stationarity in time. These efforts have been aided and complemented by numerical and physical experiments which provide important clues as to the extent to which such analytical results are robust with respect to departures from these simplifying assumptions. 

To date, the majority of our understanding of fluid turbulence is for non-relativistic, 
incompressible fluid flows described by the Navier-Stokes equation. The relativistic regime, which is
necessarily compressible, has received less attention.
Nevertheless, many applications of interest naturally require its consideration. Examples include
astrophysical fluid flows (e.g. \cite{Lynn:2014dya}), as well as applications of the fluid/gravity correspondence (see \cite{Hubeny:2011hd} and
references therein). 
This correspondence indicates that the behaviour of large black holes in asymptotically Anti-deSitter spacetimes disturbed by long wavelength perturbations can be studied by considering the relativistic hydrodynamics of a conformal fluid. 
In particular, the correspondence relates the fluid stress tensor to the 
asymptotic behaviour of the spacetime metric, as well as to intrinsic and extrinsic geometrical data of the black hole horizon.
Thus, the understanding of turbulent relativistic fluids bears relevance also to the study of gravity.

In the present work, we add to a handful of steps already taken to understand relativistic 
turbulence~\cite{Fouxon:2009rd,Eling:2009pb,Liu:2010jg,Carrasco:2012nf,Radice:2012pq,Green:2013zba,Eling:2013sna,Adams:2013vsa} by performing both analytical and numerical
analysis of such fluid flows, placing particular emphasis on two spatial dimensions.
In part, we build upon previous work by Fouxon and Oz \cite{Fouxon:2009rd}, who derived some scaling relations for relativistic hydrodynamic turbulence applicable to $d\ge 3$ spatial
dimensions. Firstly, we present some useful remarks regarding the special case of spatial dimension $d=2$, together with new scaling relations in this case in both the inverse- and direct-cascade ranges. Secondly, we describe the current state of our numerical simulations of forced turbulence on a toroidal spatial domain, with the full spacetime topology being given by $\mathcal{T}^2 \times \mathbb{R}$.

This work is organized as follows. Sec.~\eqref{sec:background} describes background information on energy scaling
and velocity correlations which are standard results in non-relativistic turbulent fluids. Sec.~\eqref{sec:relativistic}
provides a discussion of some analogous concepts in the relativistic case, and derivations of
scaling relations for this regime, with a particular attention to the dependence on dimensionality. New relativistic scaling relations will be derived
for the hydrodynamic stress-energy tensor and vorticity both in the inverse energy cascade (\ref{R1}) and in the direct enstrophy cascade (\ref{R2}), (\ref{R3}) (see also (\ref{R4})). They reduce in the non-relativistic limit to known scaling relations of incompressible fluid turbulence.
Sec.~\eqref{sec:numerics}
describes the numerical implementation employed, the initial conditions adopted as well as the statistical properties of the resulting weakly-compressible turbulent flow. We illustrate such
compressibility through Fig.~\eqref{fig:vrho-histograms}, which shows that both absolute and relative velocities are on the order of $20\%$ of the speed of sound. In this regime, we demonstrate that the term $\left\langle \rho^\prime \rho \gamma^{\prime 2} v^\prime_L \right\rangle$ in Eq.~\eqref{eq:LLLexpansion} is highly sensitive to compressive effects (see Fig.~\eqref{fig:maincorrs} and~\eqref{fig:LLLwout-spoiler}), at least insofar as it has a much wider probability distribution than its incompressible counterpart $\left\langle v^\prime_L \right\rangle$. This term might give a non-negligible contribution upon an increase in sample size since it cannot be argued to vanish by statistical symmetries, unlike its incompressible counterpart. We summarize in Sec.~\eqref{sec:summary}, and we have also included relevant information in the appendices, which we hope will prove useful for newcomers to the subject.

In this work, letters in the beginning of the alphabet $\{a$, $b$, $c$...$\}$ will denote spacetime indices, while those beginning from $\{i$, $j$, $k$...$\}$ will denote purely spatial ones. We adopt a Minkowski metric with signature $(-,+,+,+)$, and we will either denote spatial vectors with a bold symbol $\boldsymbol{r}$ or with index notation $r_i$, where appropriate. Furthermore, square brackets $\left[ . \right]$ will be used in Sec.~\eqref{sec:energyscaling} to refer to a quantity's units. Angle brackets $\left\langle . \right\rangle$ will denote ensemble averages. Finally, we use units in which the speed of light $c=1$.
%
%
\section{Background: Non-relativistic Fluid Turbulence}\label{sec:background}
The characteristics of turbulence are most cleanly studied within \emph{inertial ranges}, which are length scales far from any friction, forcing, or viscous scales. In inertial ranges, the transfer of an inviscidly conserved quantity is independent of scale. Consequently, key aspects of the analysis are often simplified. One illustration is the possibility
of using simple dimensional arguments to derive the famous Kolmogorov scaling as described in Sec.~\eqref{sec:energyscaling}.

Of particular interest for our discussion is the observation that the number of distinct inertial ranges that can exist has a dependence on dimensionality. In spatial dimensions $d>2$, if energy is being injected at a large scale $L_{f}$ and is being dissipated (e.g. by viscosity) at a small scale $L_{\nu}$, then there will be an inertial range at length scales $L$ such that $L_\nu \ll L \ll L_f$ for which the rate of energy flow between scales is independent of scale. On the other hand, for $d=2$, there exists an additional inviscidly conserved quantity called \emph{enstrophy} which gives rise to a second inertial range~\cite{Kraichnan:1967}. In what follows, we discuss some classic results in these ranges for $d=2$ which are particularly relevant for
our discussion (the interested reader may consult e.g.~\cite{Boffetta:2012} and references therein for further details).
%
\subsection{Energy scaling}\label{sec:energyscaling}
In the inertial range for energy in $d=2$, the specific kinetic energy $E = v^2/2$ transfers preferentially toward larger length scales
(since this behaviour is opposite to the $d>2$ case, this inertial range is referred to as the \emph{inverse-cascade range}). On the
other hand, the specific \emph{enstrophy}, defined by $Z = \omega^2/2$ (where $\omega = \boldsymbol{\nabla}\times\boldsymbol{v}$ is the vorticity, a pseudo-scalar quantity in $d=2$), is associated with the \emph{direct-cascade range}, so-named since it transfers preferentially toward smaller length scales. In these ranges, power-law scaling of the specific energy spectrum $E(k)$ can be obtained by dimensional analysis \cite{Boffetta:2012} as reviewed below. (Note that unless otherwise specified, energy and enstrophy are given per unit mass).

The energy spectrum has units of energy per wavenumber, or
\begin{eqnarray}
\left[ E(k)\right] = \frac{length^3}{time^2}. \label{eq:Ek1stunits}
\end{eqnarray}
Let us restrict to the energy inertial range where the rate of energy transfer through scale $k$ is given by $\epsilon \neq \epsilon (k)$. Now, assuming that the only relevant scales are $\epsilon$ and $k$, the ansatz $E(k) \propto \epsilon^p k^q$ for some $p$ and $q$ allows us to solve for the powers $p,q$ through a comparison with Eq.~\eqref{eq:Ek1stunits}:
\begin{eqnarray}
\frac{length^3}{time^2} = \left(\frac{(length^2/time^2)}{time}\right)^p \left( \frac{1}{length} \right)^q 
= \frac{length^{2p-q}}{time^{3p}} \label{eq:Ek2ndunits}.
\end{eqnarray}
This yields $p=2/3$, $q=-5/3$, which is the famous Kolmogorov scaling (see e.g.\cite{Boffetta:2012}),
\begin{eqnarray}
E(k) \propto \epsilon^{2/3} k^{-5/3}. \label{eq:fivethirds}
\end{eqnarray}
This scaling, theoretically obtained for all dimensions, has been reported in early experiments of 3D turbulence, such as in a 
jet of air under laboratory conditions \cite{Champagne:1978}, and in effectively-2D turbulence, such as in planetary atmospheres \cite{Gage:1985}, electromagnetic-layer experiments where a thin layer of electrolyte is externally forced by magnetic fields \cite{Xia:2011}, etc. Numerical experiments have also shown such behavior in, e.g.~simulations of forced, steady-state, 2D incompressible Navier-Stokes tubulence~\cite{Boffetta:2000}.
It is known to be violated, however, both numerically and experimentally in $d>2$, leading to an anomalous
scaling exponent (see the discussion in~\cite{Eling:2015mxa} and references therein e.g.~\cite{Benzi}).

In the special case of $d=2$, a further relation can be obtained which is valid in the direct-cascade range. Here, the transfer of enstrophy $\eta$ towards small scales is independent of scale. Using an analogous ansatz in this range, $E(k) \propto \eta^p k^q$, gives
\begin{equation}
\frac{length^3}{time^2} = \left(\frac{(1/time^2)}{time}\right)^p \left( \frac{1}{length} \right)^q 
= \frac{length^{-q}}{time^{3p}} \label{eq:Ekenstrophyunits}.
\end{equation}
This yields $p=2/3$ and $q=-3$, thus giving a different scaling of the energy,
\begin{equation}
E(k) \propto \eta^{2/3} k^{-3} \label{eq:minusthreelaw},
\end{equation}
for which there is a dimensionless multiplicative logarithmic correction which we will not discuss (see \cite{Boffetta:2012}). This scaling has been observed simultaneously with the $-5/3$-scaling in both 2d turbulence in a soap film \cite{Rutgers:1998}, and more tentatively in the limit of very high spatial resolution in numerical simulations \cite{Boffetta:2010}.
%
%
\subsection{Velocity structure functions}
Another classical result in the theory of turbulence is the scaling of velocity structure functions in the 2D inverse cascade range,
which highlights important statistical correlations in a turbulent flow. 
 A velocity structure function of order $n$ is a Galilean invariant, defined as
\begin{eqnarray}
\left\langle \prod_{i=1}^n (\boldsymbol{v}(\boldsymbol{r}) - \boldsymbol{v}(\boldsymbol{0})) \cdot \hat{\boldsymbol{e}}_i \right\rangle \label{eq:velstrucfunc},
\end{eqnarray}
where each $\hat{\boldsymbol{e}}_i$ is a unit vector oriented in some fixed direction with respect to the spatial separation $\boldsymbol{r}$, and the angle brackets $\left\langle . \right\rangle$ denote an ensemble average. In statistically isotropic conditions, it suffices to consider only \emph{longitudinal} and \emph{transverse} directions $\hat{\boldsymbol{e}}_L, \hat{\boldsymbol{e}}_T$, which are parallel and perpendicular to the vector $\boldsymbol{r}$, 
respectively (see \cite{Karman:1938}). 
For brevity, let us define $\delta v_\parallel (r) = (\boldsymbol{v}(\boldsymbol{r})-\boldsymbol{v}(\boldsymbol{0})) \cdot \hat{\boldsymbol{r}}$ and $\delta v_\perp (r) = (\boldsymbol{v}(\boldsymbol{r})-\boldsymbol{v}(\boldsymbol{0})) \cdot \hat{\boldsymbol{r}}_\perp$, where isotropy now serves not only to make the notion of \emph{transverse} unambiguous, but also implies these quantities depend only on the distance $r=\left| \boldsymbol{r} \right|$.

For non-relativistic, incompressible turbulent flows, one can derive scaling relations for velocity
structure functions by introducing a statistically homogeneous, isotropic, random external force.
The external force helps to establish the inertial ranges, and its statistical characteristics allow for a clean calculation (see~\cite{Bernard:1999} for the $d=2$ case).
The force has an energy injection rate $\epsilon_I$, in terms of which one finds in the inverse-cascade range $r \gg L_{forcing}$ in $d=2$ that 
\begin{eqnarray}
\left\langle \left[ \delta v_\parallel(r) \right]^3\right\rangle = 3 \left\langle \delta v_\parallel (r) \left[ \delta v_\perp (r) \right]^2 \right\rangle = \frac{3}{2} \epsilon _I r \label{eq:3/2law}.
\end{eqnarray}
If we suppose that this relation implies a scaling of the individual velocity differences as $\delta v \propto r^{1/3}$, then this immediately implies a scaling for all orders of structure functions $S_n(r)$, with any mixing between longitudinal and transverse components, given by
\begin{eqnarray}
S_n(r) \propto r^{n/3},
\label{K41}
\end{eqnarray}
provided only an even number of transverse components appear (see Appendix~\eqref{app:consequencesisotropy} for an elaboration of this point). This general scaling has been observed in various experiments \cite{Vorobieff:1999,Xia:2011,Wroblewski:2010}, as well as in forced 2D Navier-Stokes turbulence \cite{Boffetta:2000}.
Note that the scaling in Eq.~\eqref{K41} is known to be violated in all direct cascades, except for the $n=3$ structure function (see \cite{Eling:2015mxa}
and references therein, e.g.
\cite{Benzi}).
 We stress that this is by no means a complete list of references, and the interested reader should see \cite{Boffetta:2012} for a survey of previous work.
%
%
\section{Relativistic Hydrodynamic Turbulence}\label{sec:relativistic}
We now turn our attention to the case of interest, namely relativistic hydrodynamics. Let us concentrate
on the equations of motion given by the conservation of the stress-energy tensor $T_{ab}$,
\begin{eqnarray}
\partial^a T_{ab} = 0, \label{eq:eom}
\end{eqnarray}
where $a,\:b,\:c\ldots$ are spacetime indices ranging from $0\ldots d$, with $d$ the spatial dimension. Our
goal is to derive the scaling behaviour of correlations which are analogous to those found in Navier-Stokes turbulence.

\subsection{Relativistic relations I: Fouxon and Oz derivation}\label{sec:ozderivation}
For the sake of our presentation, we now
reproduce in a more detailed manner the derivation of the scaling relations presented by Fouxon and Oz \cite{Fouxon:2009rd}
for the particular case of relativistic hydrodynamics (see also \cite{Falkovich:2009mb} for compressible non-relativistic turbulence). Our notation, however, will differ: quantities evaluated at the point $\boldsymbol{r}_2$ will have a prime, while quantities evaluated at the point $\boldsymbol{r}_1$ will not.

As for Navier-Stokes turbulence, by including a random, homogeneous, and isotropic external force in the equation of motion, the inertial regime can be explored. We begin with
\begin{eqnarray}
\partial^a T_{ab} = f_b. \label{eq:eom2}
\end{eqnarray}
and assume a \emph{steady-state condition},
\begin{eqnarray}
\partial^0 \left\langle T_{0i}(t) T_{0i}^\prime(t) \right\rangle = 0, \label{eq:steadystatecondition}
\end{eqnarray}
with no sum on $i$. We stress that this condition is stronger than if we were to sum over $i$, since in the single-point limit it enforces the stationarity of average momentum $T_{0i}$ in separate directions individually, whereas summing would enforce the stationarity of the total. Acting with the derivative gives
\begin{eqnarray}
0 &=& \left\langle \partial^0 T_{0i}(t) T_{0i}^\prime(t) \right\rangle + \left\langle T_{0i}(t) \partial^0 T_{0i}^\prime(t) \right\rangle. \label{eq:steadystateexpanded}
\end{eqnarray}
Notice that interchanging the points 1 and 2 amounts to inverting the spatial coordinate axes, but this leaves the product $T_{0i}T^\prime_{0i}$ unchanged. This can be easily seen by considering a perfect fluid  $T_{ab} = (\rho+p)u_a u_b + p\eta_{ab}$,
expressing $\boldsymbol{u} = \gamma (1, v_i)$, where $v_i$ is the spatial velocity, and realizing that the switch changes the sign of each
$v_i$ but the product $v v'$ remains unchanged. We can therefore switch points 1 and 2 in the first term of 
Eq.~\eqref{eq:steadystateexpanded} without consequence, which combines the two terms to give
\begin{eqnarray}
0  &=&  \left\langle T_{0i}(t) \partial^0 T_{0i}^\prime(t) \right\rangle, \nonumber \\
&=& \left\langle T_{0i}(t) [ -\partial^{\prime j} T_{ij}^\prime(t) + f_i^\prime(t) ] \right\rangle, \label{eq:timederivofcorr}
\end{eqnarray}
where we have used Eq.~\eqref{eq:eom2} to replace the time derivative, and a sum on $j$ is understood. It is important to stress that the spatial derivative here is with respect to the coordinates of point 2 (we have denoted this with a prime, $\partial^\prime$). This means that it views functions of $\boldsymbol{r}_1$ as constant, and so can be brought out of the ensemble average.
Thus,
\begin{eqnarray}
\partial^{\prime j} \left\langle T_{0i}(t)  T_{ij}^\prime(t)  \right\rangle = \left\langle T_{0i}(t) f_i^\prime (t) \right\rangle. \nonumber
\end{eqnarray}
Now, notice that homogeneity implies that these averaged quantities are functions of the separation $\boldsymbol{r} \equiv  \boldsymbol{r}_2 - \boldsymbol{r}_1 $ only, so $\partial^{\prime j} \equiv \partial/\partial (r_2)_j = \partial/\partial r_j \equiv \partial^j$ when acting on them. This gives,
\begin{eqnarray}
\partial^j \left\langle T_{0i}(t)  T_{ij}^\prime(t)  \right\rangle = \left\langle T_{0i}(t) f_i^\prime (t) \right\rangle. \label{eq:beforeintegration}
\end{eqnarray}
Assuming that $r \ll L_f \equiv $ the correlation length of the forcing allows the approximation $\left\langle T_{0i}(t) f_i^\prime (t) \right\rangle \approx \left\langle T_{0i}(t) f_i (t) \right\rangle \equiv \epsilon_i$, which is now constant with respect to $r$. Using the fact that the left-hand side is a gradient and $\left\langle T_{0i}(t) T^\prime_{ij}(t) \right\rangle$ is isotropic (not a function of angle), Eq.~\eqref{eq:beforeintegration} can be integrated over a disc using the divergence theorem. This yields
\begin{eqnarray}
\left\langle T_{0i}(t)  T_{ij}^\prime(t)  \right\rangle = \frac{\epsilon_i r_j}{d}. \label{eq:ozresult}
\end{eqnarray}
This completes the derivation. Assuming instead that $r \gg L_f$ does not allow one to integrate Eq.~\eqref{eq:beforeintegration} without more information, as the result would depend on the details of the forcing at all scales up to $r$. Also, note that if one wished to enforce that $f_i$ is divergence-free, then $f_i$ would only be isotropic in the sense that $\left\langle \hat{f}_i (\boldsymbol{k}) \hat{f}^i (\boldsymbol{k}) \right\rangle$ is a function of $k$ only, so one should sum over $i$ in Eq.~\eqref{eq:ozresult} in that case.
%
%
%
%
\subsection{Relativistic relations II: the case of d=2}\label{sec:alternativederivation}
In this section we concentrate on the behaviour of $\left\langle T_{0i}(t)  T_{ij}^\prime(t)  \right\rangle$ for the special case of $d=2$ in the {\em inverse-cascade range}, as well as an additional correlation function in the direct-cascade range which involves a quantity resembling vorticity.

A special treatment of $d=2$ is required, since 
whether the steady-state condition~\eqref{eq:steadystatecondition} is appropriate depends upon whether the energy injected by the external force can be removed. For $d>2$, since injected energy transfers to small scales, it will encounter the viscous scale and be dissipated. There is evidence that this behavior persists even for arbitrarily small viscosity, and is known as the \emph{energy dissipation anomaly} (see~\cite{Frisch:1996}). This can be understood heuristically as a result of the direct cascade of energy; a finite viscosity, no matter how small, will produce strong energy dissipation below the viscous scale, and the direct cascade of energy guarantees that this scale will eventually be encountered. Mathematically, this can be understood in the incompressible case as a result of the unboundedness of enstrophy. One can derive the energy balance equation with no forcing or friction~\cite{Frisch:1996},
\begin{eqnarray}
\frac{dE}{dt} = -2 \nu \Omega, \label{eq:energybalance}
\end{eqnarray}
where $E \equiv \left\langle v^2/2 \right\rangle$ is the \emph{mean energy} and $\Omega \equiv \left\langle \omega^2/2 \right\rangle$ is the \emph{mean enstrophy}. If $\Omega$ can become comparatively large as the viscosity $\nu$ becomes small, then the right-hand side of Eq.~\eqref{eq:energybalance} can remain non-zero. The balance equation for $\Omega$ contains a source term which is due to \emph{vortex stretching}, preventing one from bounding its growth. Thus, if the anomalous energy dissipation persists in the relativistic regime for $d>2$, then the energy injected by the external force can be removed and the steady-state condition \eqref{eq:steadystatecondition} is appropriate.

On the other hand, the situation is different when $d=2$. Vortex stretching is absent in this case, which means no source term appears in the enstrophy balance equation, thus being given by~\cite{Boffetta:2012}
\begin{eqnarray}
\frac{d\Omega}{dt} = -2 \nu P, \label{eq:enstrophybalance}
\end{eqnarray}
where $P = \left\langle \left| \boldsymbol{\nabla}\omega \right|^2/2 \right\rangle$ is the \emph{mean palinstrophy}, and where we have again restricted to the incompressible case with no forcing or friction. Eq.~\eqref{eq:enstrophybalance} says that the mean enstrophy is dissipated in time, which means it is bounded from above. It follows that the energy dissipation vanishes in the limit of zero viscosity, since the enstrophy cannot grow comparatively. If this fact remains true in the relativistic regime, then the steady-state condition~\eqref{eq:steadystatecondition} is inappropriate for $d=2$ in the inviscid limit. Even without taking that limit, when the forcing and viscous scales are sufficiently separated, one would expect energy dissipation to be small (and to remain small over time, due to the inversely-cascading energy), thus easily failing to balance the injection of energy.

Thus, in the next section we present an alternative derivation which gives rise to different scaling 
behaviour of the same correlation functions. (It is worth mentioning large-scale energy may 
transfer towards the viscous scale through the formation of shocks or large gradients, 
where it would be dissipated. 
This might allow the steady-state condition~\eqref{eq:steadystatecondition} to hold in $d=2$.) 

Lastly, in $d=2$ the balance equation for the palinstrophy $P$ reads
\begin{eqnarray}
\frac{dP}{dt} = -\nu \left\langle \left( \nabla^2 \omega\right)^2 \right\rangle - \left\langle \left( \partial_i \omega \right) \left( \partial_j \omega \right) \left( \partial^j u^i \right) \right\rangle ,
\end{eqnarray}
which has a source term of indefinite sign. This means that the palinstrophy $P$ cannot be bounded from above, so there may be an \emph{enstrophy dissipation anomaly} in $d=2$~\cite{Boffetta:2012,Bernard:1999} if $P$ can become large enough that the right-hand side of Eq.~\eqref{eq:enstrophybalance} remains non-zero for arbitrarily small $\nu$. This will allow us to derive new correlation functions in the relativistic case by considering a different steady-state condition involving quantities that resemble vorticity.

\subsubsection{Scaling in the inverse-cascade range}\label{sec:inversecascadescaling}
We now derive a relativistic scaling relation in the inverse-cascade range by adapting a strategy used in the incompressible Navier-Stokes case~\cite{Bernard:1999}.
Let us begin by defining a quantity $\epsilon$  by
\begin{eqnarray}
\epsilon \equiv  \partial^0 \left\langle \frac{T_{0i}T_0^i}{2} \right\rangle , \label{eq:2ndsinglepointlimit}
\end{eqnarray}
where we are summing over $i$ this time. What follows does not require $\epsilon$ to be independent of time. Consider a new form of stationarity, weaker than Eq.~\eqref{eq:steadystatecondition}, which is
consistent with a lack of removal of energy,
\begin{eqnarray}
0= \partial^0 \left\langle ( T_{0i}^\prime - T_{0i} )( T_0^{\prime i}- T_0^i ) \right\rangle = \partial^0 \left\langle {T_{0i}^\prime} T_0^{\prime i} + T_{0i} T_0^i - 2 T_{0i} T_0^{\prime i} \right\rangle. \label{weakerrelation1}
\end{eqnarray}
Notice that expression (\ref{weakerrelation1}) reduces in the Newtonian limit to the stationarity of a second-order velocity structure function.
Now, homogeneity implies $\left\langle{T_{0i}^\prime} T_0^{\prime i} \right\rangle = \left\langle T_{0i} T_0^i \right\rangle$, and recall that we have already evaluated the third term on the right-hand side of this equation 
in Eq.~\eqref{eq:timederivofcorr}. Upon replacement, one obtains
\begin{eqnarray}
0= \partial^0 \left\langle T_{0i} T_0^i \right\rangle - 2\left\langle T_{0i} f^{\prime i} \right\rangle + 2\partial^j \left\langle T_{0i} T_{j}^{\prime i} \right\rangle, \nonumber
\end{eqnarray}
or, using the definition Eq.~\eqref{eq:2ndsinglepointlimit},
\begin{eqnarray}
\partial^j \left\langle T_{0i} T_{j}^{\prime i} \right\rangle &=& \left\langle T_{0i}f^{\prime i} \right\rangle - \epsilon. \label{eq:beforeint_alternative}
\end{eqnarray}

At this point, we must relate $\left\langle T_{0i} f^{\prime i} \right\rangle$ to our choice of external force. 
We adopt a divergence-free homogeneous Gaussian random field with zero mean, characterized by its two-point correlation function (see e.g.~\cite{cannonthesis}),
\begin{eqnarray}
\left\langle f^\prime_i (t^\prime) f_j (t) \right\rangle = F_{ij}(\boldsymbol{r}) \delta(t^\prime - t), \label{eq:beforeNovikov}
\end{eqnarray}
such that $F_{ij}$ decays rapidly beyond the forcing scale $L_f$. We impose isotropy in the sense that $\text{tr}F \equiv F_i^i$ is a function of $r$ only and,
as shown in Appendix~\eqref{app:Novikov}, $\text{tr}F = 2 \left\langle T_{0i}f^{\prime i} \right\rangle$, which gives us greater control over this term. Furthermore, since $f_i$ is divergence-free, $\text{tr}F = \partial^k \Theta_{k}$ for some appropriate $\Theta_{k}$. This can be seen by first noting that $f_i$, if divergence-free, can be written in terms of a stream function $\psi$ as $f_i = \epsilon_{ij}\partial^j \psi$. Thus,
\begin{eqnarray}
\left\langle f_i f^{\prime i} \right\rangle &=& \left\langle ( \epsilon_{ik}\partial^{k} \psi ) (\epsilon^i_{n} \partial^{\prime n} \psi^\prime ) \right\rangle \nonumber\\
&=& \partial^k \left\langle ( \epsilon_{ik} \psi ) (\epsilon^i_{n} \partial^{\prime n} \psi^\prime ) \right\rangle.
\end{eqnarray}
We can then relate this to $\text{tr}F$ by integrating Eq.~\eqref{eq:beforeNovikov} with respect to time to eliminate the $\delta$-function, then defining $\Theta_{k} \equiv \int d\tau \left\langle ( \epsilon_{ik} \psi ) (\epsilon^i_{n}\partial^{\prime n} \psi^\prime ) \right\rangle$, thereby showing $\text{tr}F = \partial^k \Theta_{k}$ by construction.

Our expression Eq.~\eqref{eq:beforeint_alternative} thus becomes
\begin{eqnarray}
\partial^j \left\langle T_{0i} T_{j}^{\prime i} \right\rangle = \frac{1}{2} \partial^j \Theta_{j} - \epsilon, \label{eq:beforeint_valideverywhere}
\end{eqnarray}
which, under the isotropic conditions assumed, integrates to 
\begin{eqnarray}
\left\langle T_{0i} T_{j}^{\prime i} \right\rangle = \frac{1}{2} \left(\Theta_{j} - \epsilon r_j \right),  \label{eq:afterint_valideverywhere}
\end{eqnarray}
a result which holds for all $r$. For the inverse-cascade range, this result further simplifies since $\Theta_{j}$ is negligible there. To see this, first note that for $j=T$ the transverse direction, $\left\langle T_{0i} T_{j}^{\prime i} \right\rangle$ vanishes by isotropy and $r_j$ vanishes by definition, so $\Theta_{T}$ must also. For $\Theta_{L}$, recall that we have
 stipulated that $\text{tr}F$ decays rapidly beyond the forcing scale $L_f$. Thus, integration of $\text{tr}F$ over a disc of radius $r$ will approach a constant as $r$ exceeds the forcing scale $L_f$, whereas applying the divergence theorem yields
\begin{eqnarray}
\int_{disc} \text{tr}F dA &=& \int_{disc} \partial^j \Theta_{j} dA \nonumber\\
&=& \int_{\partial (disc)} \Theta_{j} \hat{r}^j r d\theta \nonumber\\
&=& 2 \pi r \Theta_{L}.
\end{eqnarray}
Thus the longitudinal component $\Theta_{L}$ decays at least as quickly as $1/r$, becoming negligible at large distance. 
Consequently,
\begin{eqnarray}
\left\langle T_{0i} T_{j}^{\prime i} \right\rangle = - \frac{\epsilon r_j}{2} \label{eq:afterint_inversecascade}
\end{eqnarray}
in the inverse-cascade range, where we have neglected the subleading term. 
Notice that this result has the {\em opposite sign} with respect to the $d>2$ case, which in the incompressible limit is known to reflect the inverse cascade of energy. As a word of caution, note that this scaling is usually presented as positive since the points $\boldsymbol{r}_2$ and $\boldsymbol{r}_1$ are switched. In other words, Eq.~\eqref{eq:afterint_inversecascade} is equivalent to
\begin{eqnarray}
\left\langle T^\prime_{0i} T_{j}^{i} \right\rangle = + \frac{\epsilon r_j}{2}.
\label{R1}
\end{eqnarray}
Thus, when comparing the overall signs in Eqs.~\eqref{eq:ozresult} and~\eqref{eq:afterint_inversecascade} with the literature, one should be mindful of this point.

It is interesting to note that in the incompressible limit, the constant $\epsilon = \left\langle v_i f^i \right\rangle$ is the lowest order term in the Taylor series of $\left\langle v_i f^{\prime i} \right\rangle$~\cite{Bernard:1999}. Thus, in the short-distance limit, the first non-zero term in the Taylor series of $\Theta_{j} - \epsilon r_j$ is proportional to $r^2 r_j$. This gives the cubic scaling of the third order velocity correlation familiar from the statistical theory of incompressible turbulence. However, incompressibility plays a crucial role in this result, so we cannot make a similar inference in the relativistic case without additional assumptions.

Finally, had we used the slightly weaker steady-state condition that $\partial^0 \left\langle(T_{0i}^\prime - T_{0i}) (T_{0}^{\prime i} - T_0^i) \right\rangle = constant$ not necessarily zero, we would clearly still obtain linear scaling in the inverse-cascade range, although with a different proportionality constant. This weaker assumption might hold on a periodic 2D spatial domain, such as a torus, in the absence of any removal of energy. However, since energy would cascade towards the longest available length scale, anisotropy would grow as energy condensates into the lowest mode (see Appendix~\eqref{sec:energycondensate} for a numerical simulation of this scenario). Thus, the linear scaling obtained here would be expected to hold only in the intermediate stage when the flow is still isotropic.

\subsubsection{Scaling in the direct-cascade range}\label{sec:directcascadescaling}
In the incompressible, non-relativistic case, the statistics in the direct-cascade range can be cleanly
studied using a steady-state condition of correlations involving the vorticity. A similar strategy can be adopted
here, although subtleties arise with regard to the precise expression of vorticity adopted. In what
follows we describe what we consider the most straightforward path and refer to Appendix~\ref{sec:vorticityappendix} for a related option.
First, consider the spatial component of Eq.~\eqref{eq:eom2},
\begin{eqnarray}
\partial^0 T_{0i} + \partial^j T_{ij} = f_i \label{eq:spatialeom},
\end{eqnarray}
and apply the 2-dimensional curl to obtain
\begin{eqnarray}
\partial^0 \left( \epsilon^{ik} \partial_k T_{0i}\right) + \partial^j \left( \epsilon^{ik} \partial_k T_{ij} \right) = \left( \epsilon^{ik} \partial_k f_i \right)\label{eq:pseudovorticity_eom_before}.
\end{eqnarray}
The incompressible limit of this is the standard equation for vorticity. However,
it is interesting to note that Eq.~\eqref{eq:pseudovorticity_eom_before} does not describe what is normally regarded as the relativistic vorticity, even though it has the same incompressible limit.~(We describe the behaviour of the relativistic vorticity in Appendix~\ref{sec:vorticityappendix}, together with a mention of the subtleties related to deriving scaling relations with it.) We identify the right-hand side of Eq.~\eqref{eq:pseudovorticity_eom} as the curl of the external force, which we will denote as $\mathcal{F}$. For brevity, let us also define the first two quantities in brackets as $\omega = \epsilon^{ik}\partial_k T_{0i}$ and $\bar{\omega}_j = \epsilon^{ik}\partial_k T_{ij}$, giving the suggestive expression
\begin{eqnarray}
\partial^0 \omega + \partial^j \bar{\omega}_j = \mathcal{F} \label{eq:pseudovorticity_eom}.
\end{eqnarray}
We may now multiply this expression by $\omega^\prime$ and take the ensemble average, which gives,
\begin{eqnarray}
\partial^0 \left\langle \frac{\omega\omega^\prime}{2} \right\rangle + \frac{\partial \left\langle \bar{\omega}_j \omega^\prime \right\rangle}{\partial (r_1)_j} = \left\langle \mathcal{F}\omega^\prime \right\rangle \label{eq:pseudoenstrophy_eom_explicit}.
\end{eqnarray}
We have explicitly shown that the spatial derivative is with respect to the point $\boldsymbol{r}_1$. It acts on a correlation which, by the assumption of homogeneity, is a function of separation $\boldsymbol{r}=\boldsymbol{r}_2-\boldsymbol{r}_1$ only. Thus, we can change $\partial/\partial(r_1)_j$ to $-\partial/\partial r_j$, which herein we write simply as $\partial^j$, thus giving 
\begin{eqnarray}
\partial^0 \left\langle \frac{\omega\omega^\prime}{2} \right\rangle - \partial^j \left\langle \bar{\omega}_j \omega^\prime \right\rangle = \left\langle \mathcal{F}\omega^\prime \right\rangle \label{eq:pseudoenstrophy_eom}.
\end{eqnarray}
 Assuming the existence of a dissipation anomaly for the quantity $\left\langle \omega^2/2 \right\rangle$, which would balance the injection from the external force, we can impose the steady-state condition
\begin{eqnarray}
\partial^0\left\langle \frac{\omega \omega^\prime}{2} \right\rangle = 0 \label{eq:pseudoenstrophy_steadystate}
\end{eqnarray}
even for arbitrarily small viscosity. Thus Eq.~\eqref{eq:pseudoenstrophy_eom} yields
\begin{eqnarray}
\partial^j \left\langle \bar{\omega}_j \omega^\prime \right\rangle = -\left\langle \mathcal{F}\omega^\prime \right\rangle \label{eq:pseudovort_beforeint}.
\end{eqnarray}
In the direct-cascade range $r\ll L_f$, $\left\langle \mathcal{F}\omega^\prime \right\rangle \approx \left\langle \mathcal{F}\omega \right\rangle \equiv \varepsilon$, which allows us to integrate Eq.~\eqref{eq:pseudovort_beforeint} using isotropy, obtaining
\begin{eqnarray}
\left\langle \bar{\omega}_j \omega^\prime \right\rangle = -\frac{\varepsilon}{2} r_j \label{eq:relvortcorr}.
\label{R2}
\end{eqnarray}

Summarizing so far, we find that $\left\langle T_{0i}T_{j}^{\prime i}\right\rangle$ scales linearly in the inverse-cascade range with the opposite sign relative to the $d>2$ case, and its linear scaling in the inverse-cascade range ought to be robust with respect to the background topology, subject to the assumption of isotropy. Furthermore, we found that $\left\langle \bar{\omega}_j \omega^\prime \right\rangle$ scales linearly in the direct-cascade range. 

Finally, it is possible to integrate Eq.~\eqref{eq:relvortcorr} twice more to obtain a cubic scaling of $\left\langle T_{0T} T_{LT}^{\prime} \right\rangle$ in the direct-cascade range, but through this procedure one obtains no information about the purely longitudinal correlation $\left\langle T_{0L} T_{LL}^{\prime} \right\rangle$. To see this, begin by writing the left-hand side of Eq.~\eqref{eq:relvortcorr} with $\bar{\omega}_j$ and $\omega^{\prime}$ appearing explicitly in terms of the stress-energy tensor,
\begin{eqnarray}
\left\langle \bar{\omega}_j \omega^\prime \right\rangle &=& \left\langle \left( \epsilon^{mn}\frac{\partial T_{mj}}{\partial(r_1)^n} \right) \left( \epsilon^{ik}\frac{\partial T^{\prime}_{0i}}{\partial (r_2)^k} \right) \right\rangle \nonumber\\
&=& \frac{\partial}{\partial(r_1)^n} \frac{\partial}{\partial (r_2)^k} \left\langle \epsilon^{mn}\epsilon^{ik} T_{mj} T^{\prime}_{0i} \right\rangle \nonumber\\
&=& -\partial_n \partial_k \left\langle \epsilon^{mn}\epsilon^{ik} T_{mj} T^{\prime}_{0i} \right\rangle,
\end{eqnarray}
where we have again used $\partial/\partial (r_2)^i = - \partial/\partial (r_1)^i = \partial/\partial r_i \equiv \partial_i$ in the last line, which is true when the derivative acts on functions of the separation $\boldsymbol{r}=\boldsymbol{r}_2-\boldsymbol{r}_1$ only. For cleanliness, define $A^n_j \equiv \partial_k \left\langle \epsilon^{mn}\epsilon^{ik} T_{mj} T^{\prime}_{0i} \right\rangle$, so that Eq.~\eqref{eq:relvortcorr} now reads
\begin{eqnarray}
\partial_n A^n_j = \frac{\varepsilon}{2} r_j.
\end{eqnarray}
We wish to integrate this over a disc using the divergence theorem, so let us obtain a scalar equation by projecting the $j$-index onto the longitudinal direction $L$, giving
\begin{eqnarray}
\partial_n A^n_L = \frac{\varepsilon}{2}r.
\end{eqnarray}
Integration over a disc, assuming isotropy so that $A^L_L \neq A^L_L(\theta)$, yields
\begin{eqnarray}
A^L_L \equiv \partial_k \left\langle \epsilon^{mL}\epsilon^{ik} T_{mL} T^{\prime}_{0i} \right\rangle = \frac{\varepsilon}{6}r^2.
\end{eqnarray}
A further application of the divergence theorem yields
\begin{eqnarray}
\left\langle \epsilon^{mL}\epsilon^{iL} T_{mL} T^{\prime}_{0i} \right\rangle = \frac{\varepsilon}{24}r^3.
\end{eqnarray}
Using the identity $\epsilon^{mL}\epsilon^{iL} = \delta^{mi}\delta^{LL} - \delta^{mL}\delta^{iL}$, one obtains the final result,
\begin{eqnarray}
\left\langle T_{TL} T^{\prime}_{0T} \right\rangle = \frac{\varepsilon}{24}r^3,
\label{R3}
\end{eqnarray}
valid in the direct-cascade range.
%
%
\section{Implementation details}\label{sec:numerics}
In order to test the derived scaling relations, we numerically implement
the relativistic hydrodynamical equations subjected to an external force
with suitable statistical properties. We then extract relevant quantities from the numerical
solution, as described below. In what
follows we provide details of our implementation.

\subsection{Flux-conservative formulation}
For convenience we express the equations of motion in 
flux-conservative form. In the absence of driving-sources, this helps to ensure energy-momentum
conservation at the discrete level. As discussed in \cite{Calabrese:2001kj,gustafsson1995time},
the combination of discrete operators satisfying {\em summation by parts} together with a Runge-Kutta integrator of third order guarantees an energy conserving scheme. Eq.~\eqref{eq:eom2} 
gives two expressions, already in the desired form,
\begin{eqnarray}
\partial^0 T_{00} + \partial^i T_{i0} = 0, \label{eq:energycons}\\
\partial^0 T_{0i} + \partial^j T_{ij} = f_i, \label{eq:momentumcons}
\end{eqnarray}
where $i$ and $j$ are spatial indices. These equations fully determine the system in the ultra-relativistic
regime where the conservation of particle number becomes irrelevant at the classical level.
We take $\{ T_{00},T_{0i} \}$ to be our set of {\em conservative variables}, and evolve them directly. Using a perfect fluid with the conformal equation of state $p=\rho/2$, the equations of motion become
\begin{eqnarray}
\begin{aligned}
\partial^0 \left(\frac{3}{2}\rho\gamma^2 - \frac{1}{2}\rho\right) + \partial^i \left(\frac{3}{2}\rho\gamma^2 v_i\right) &=0 \nonumber\\
\partial^0 \left(\frac{3}{2}\rho\gamma^2 v_i\right) + \partial^j \left(\frac{3}{2}\rho\gamma^2 v_i v_j +  \frac{1}{2}\rho\delta_{ij}\right) &=f_i, \nonumber
\end{aligned}
\end{eqnarray}
where we have used $u_a = \gamma (-1,\vec{v})$. We define our conservative variables as $D\equiv(\rho/2) (3\gamma^2-1)$, $S_i \equiv (3/2)\rho\gamma^2 v_i$ and our primitive variables as $(\rho, v_i)$ for $i=1,2$. Note that the second equation provides the time-evolution of $S_i$, which then sources the first equation for the time-evolution of $D$. The forcing function $f_i$, which is completely spatial, is described in Appendix (\ref{sec:forcingfunction}).

The transformation from conservative variables $(D,S_i)$ to primitive variables $(\rho,v_i)$ is given by
\begin{eqnarray}
\rho = \frac{2D}{3\gamma^2 -1},\; v_i = \frac{2 S_i}{3\gamma^2\rho}, \label{eq:contoprim}
\end{eqnarray}
Solving for the Lorentz factor in terms of the conservative variables amounts to solving a quadratic equation for $\gamma^2$. The presence of $\rho$ in the denominator presents a potential problem in the recovery of $v_i$ when $\rho=0$. In general applications, this technical issue can be circumvented by artificially maintaining a non-zero \emph{floor} or \emph{atmosphere} for $\rho$, small enough so as not to affect the dynamics appreciably. However, in our simulations the density never reaches zero, so this mechanism is never invoked. 
%
\subsection{Spatio-temporal reduction of the ensemble average $\left\langle . \right\rangle$} \label{sec:spatiotempreduc}
It is often impractical to calculate $\left\langle . \right\rangle$ as an ensemble average. In practice, one exploits statistical symmetries and the assumption of ergodicity to reduce $\left\langle . \right\rangle$ to a spatial or temporal average. For instance, for a statistically homogeneous and isotropic flow, one computes $\left\langle . \right\rangle$ as an average over pairs of points with a scalar separation $r = |\vec{\boldsymbol{r}}|$. Further details on the mathematical subtleties involved in doing these reductions rigorously are given in \cite{Frisch:1996}, and will not be discussed here.

Taking Eq.~\eqref{eq:velstrucfunc} as an example, the homogeneous and isotropic averaging process means that all quantities in the product are projected onto directions defined relative to the separation vector $\boldsymbol{r}$. Thus, when computing the average spatially on a numerical grid, although $\boldsymbol{r}$ itself may vary in direction from term to term, the relative directions between $\boldsymbol{r}$ and the projection directions must remain the same. 

With this in mind, we understand that the spatial indices in Eq.~\eqref{eq:ozresult} stand for projections onto those directions. This implies that when the $j$-th direction, $\hat{\boldsymbol{j}}$, is perpendicular with $\vec{\boldsymbol{r}}$, the correlation in Eq.~\eqref{eq:ozresult} vanishes since $r_j = \vec{\boldsymbol{r}}\cdot \hat{\boldsymbol{j}} = 0$. This is a consequence of isotropy, and we elaborate on it in Appendix (\ref{app:consequencesisotropy}). 
%
%
\subsection{Numerical experiments}
Our simulations take place on a torus and, as mentioned, unless some 
some form of large-scale extraction of energy is employed, energy will build up in the longest mode. To address
this issue, we adopt a convenient approach by augmenting Eq.~\eqref{eq:momentumcons} with a linear `friction' term $-\alpha T_{0i}$ on the right-hand side\footnote{Alternatively,
one can in principle remove this energy build-up through a suitable analysis as
described in Appendix~\ref{sec:energycondensate}. We have however found the approach employing the
friction term is more straightforward.}, giving
\begin{eqnarray}
\partial^0 T_{0i} + \partial^j T_{ij} = f_i -\alpha T_{0i}. \label{eq:momentumcons_withfriction}
\end{eqnarray}
The friction term causes the system to evolve towards an approximately constant total energy. For sufficiently small $\alpha > 0$, this final state exhibits inertial range scaling in the Newtonian spectral energy $E(k)$, and thereby exhibits fully developed turbulent behaviour. By dimensional analysis, $\alpha$ can be related~\cite{Boffetta:2012} to the energy injection rate $\epsilon_I$ and the friction length-scale $L_\alpha$ through
\begin{eqnarray}
\alpha \sim \left( \frac{\epsilon_I}{L_\alpha^2}\right)^{1/3} \label{eq:frictionscale}.
\end{eqnarray}
For a given $\epsilon_I$, we choose $\alpha$ such that the friction length scale $L_\alpha$ is a few times smaller than the spatial extent of the domain. We next describe the setup of our simulation of fully-developed, steady-state turbulence, described by Eq.~\eqref{eq:momentumcons_withfriction}.

\begin{figure}[h!]
\centering
\includegraphics[width=1.\textwidth]{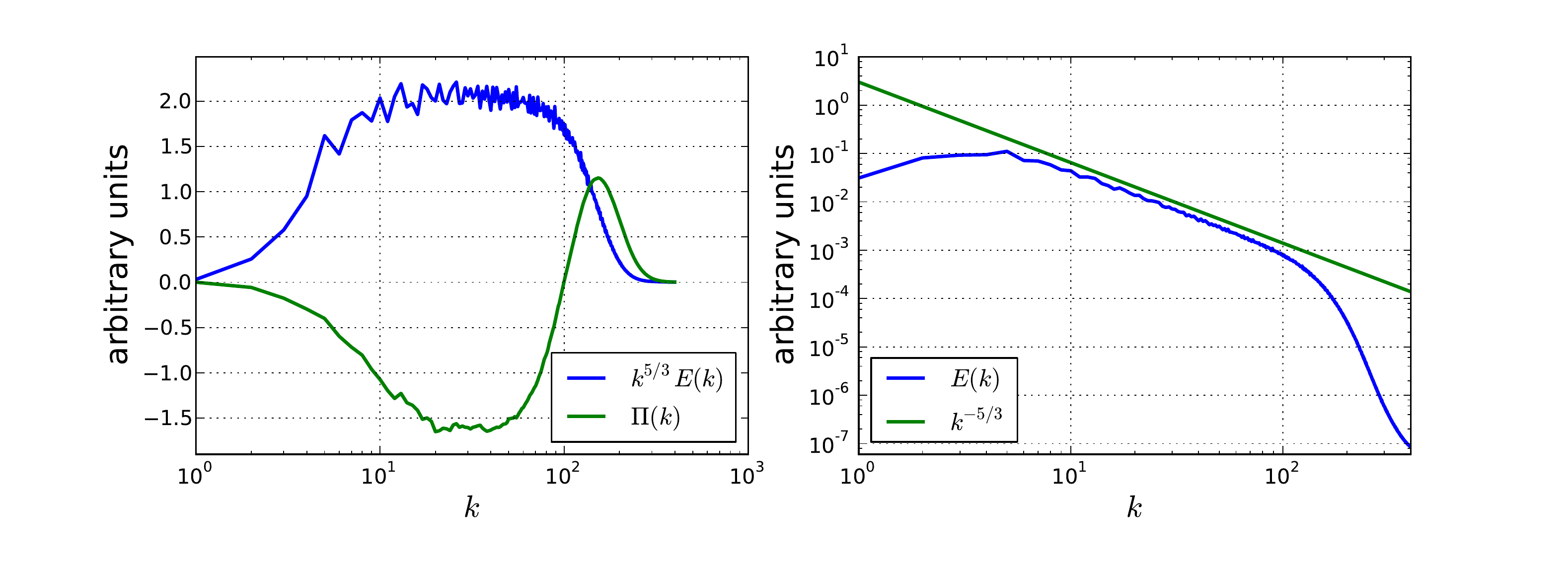}
\caption{Left: Isotropic Newtonian spectral energy $E(k)$ and energy flux $\Pi(k)$ averaged over $200$ flow realizations. $E(k)$ is compensated by the inverse Kolmogorov power law $k^{5/3}$, and both quantities are scaled to a convenient, comparable magnitude for the purposes of presentation. Note the semi-log scale. The energy flux $\Pi(k)$ crosses zero at $k=100$, indicating the injection of energy there, while it takes on negative values for $k<100$, indicating the inverse-cascade of energy. Right: For ease of visual comparison, we plot $E(k)$ in a manner that is common in previous work, eg. \cite{Boffetta:2010}.}\label{fig:E(k)}
\end{figure}

The initial conditions adopted are $\{\rho=1$, $v_i=0\}$, and the uniform spatial grid has $N^2 = 800^2$ points (which admits the Nyquist wavenumber $k_{\text{max}} = 400$, expressed in grid units; one can convert to real units via $2\pi k_{\text{max}}/L$) and periodic boundary conditions are imposed. For concreteness,
all reported times will be given in multiples of the light-crossing time $t_{LC} = L/c$, with $L \equiv$ the size of the box, which we set numerically to $10$. The random external force employed is described in Appendix~\eqref{sec:forcingfunction}, and with its strength controlled by the parameter $\Psi (0)=3\times 10^{-5}$ we find a suitable value of the friction strength to be $\alpha=1.8\times 10^{-2}$, producing a large-scale energy cutoff around $k=5$, as shown in Fig.~\eqref{fig:E(k)}.

\begin{figure}[h!]
\centering
\includegraphics[width=\textwidth]{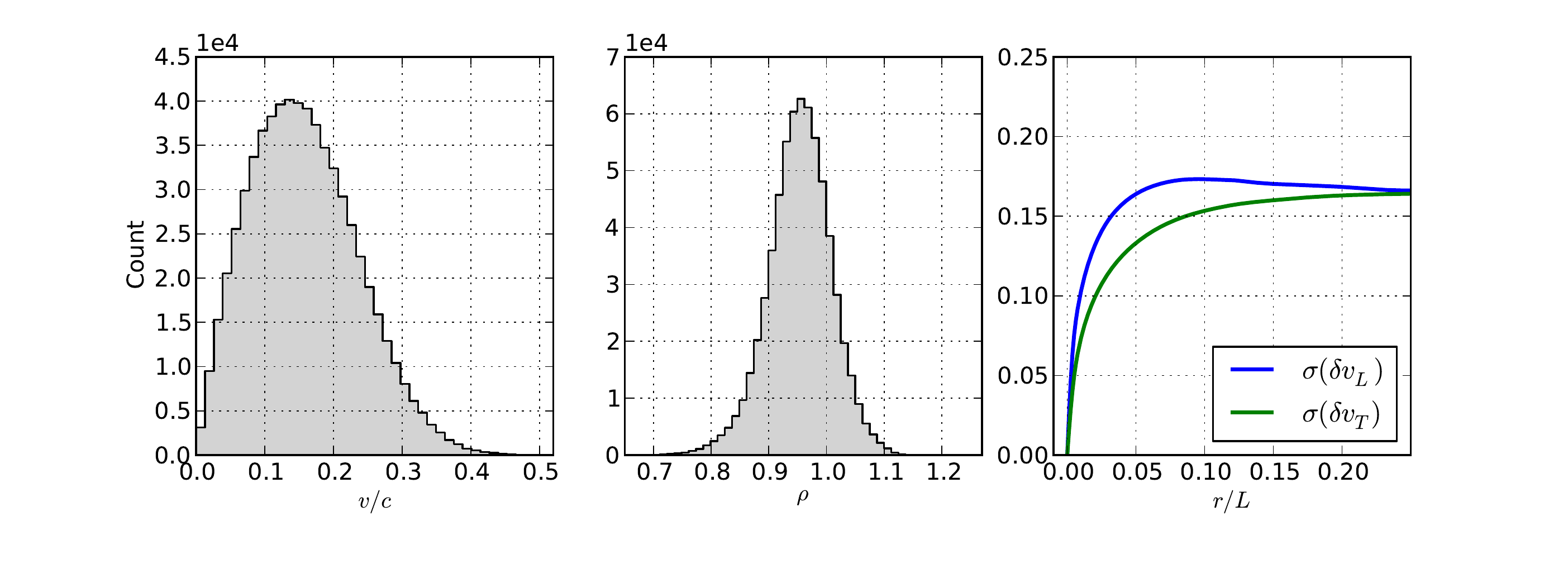}
\caption{Velocity (left) and density (middle) distributions for a single representative realization of the flow. The velocity peaks at $v= 0.14c$ and the highest velocity is $= 0.52c$. The density peaks at $\rho= 0.97$, and has a standard deviation of $0.055$. Right: the standard deviation of the longitudinal and transverse velocity differences $\delta v_L$, $\delta v_T$, as a function of separation $r/L$, as drawn from 10 realizations of the flow. The distributions of these velocity differences are roughly Gaussian with zero mean. Note the overall magnitude of $\sim 0.16$ as compared with the sound speed $0.71$.)}\label{fig:vrho-histograms}
\end{figure}

\begin{figure}[h!]
\centering
\includegraphics[width=.5\textwidth]{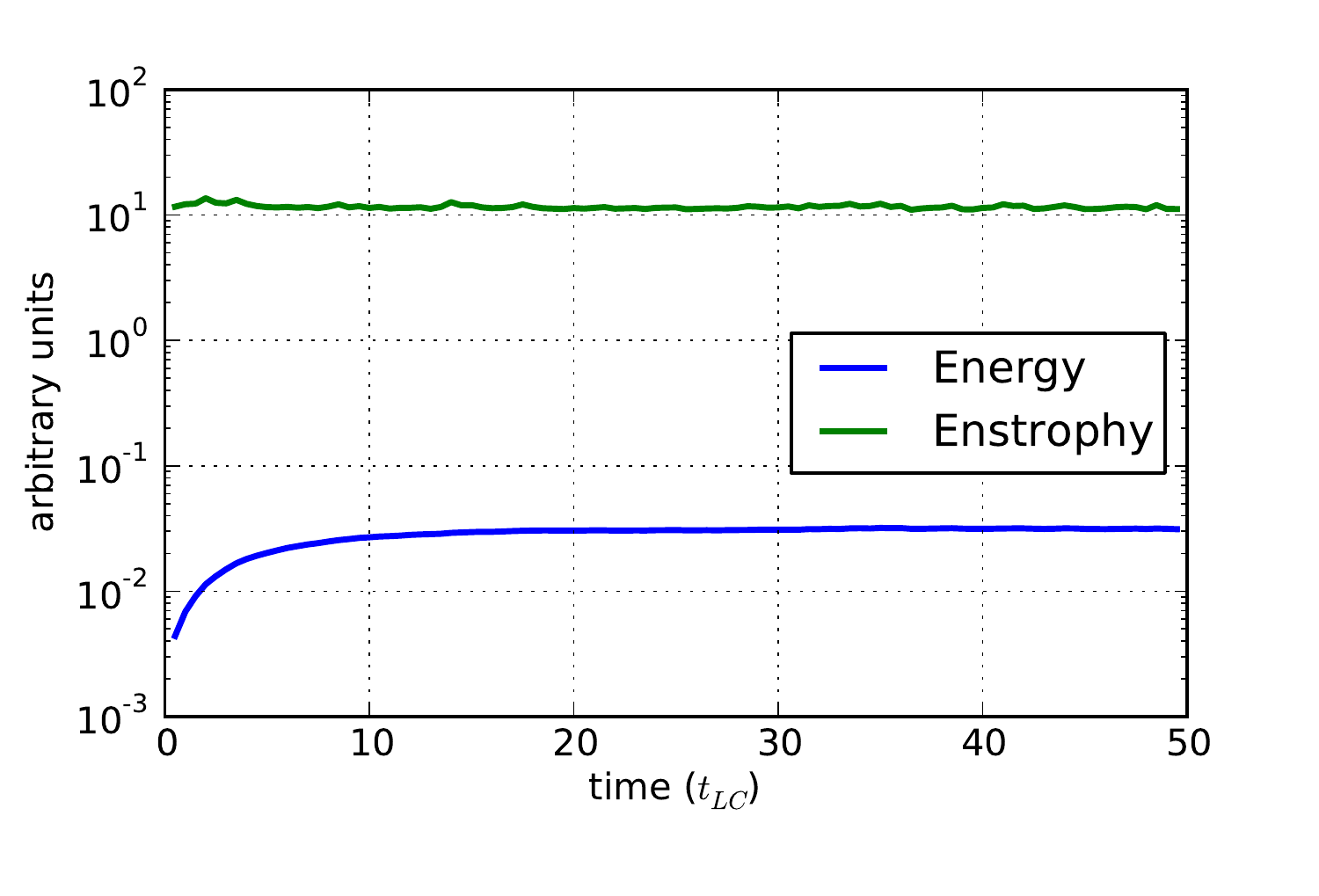}
\caption{Total Newtonian kinetic energy and total Newtonian enstrophy displayed as functions of time, measured in multiples of the light-crossing time $t_{LC}$ of the box. Plateaus occur quickly, indicating a statistically steady-state. Analysis is performed at $t=25t_{LC}$ for each run.}\label{fig:Et-and-Wt}

\end{figure}

Fig.~\eqref{fig:vrho-histograms} displays the velocity and density distributions at a representative time when the flow appears statistically stationary, as well as the standard deviation of the velocity differences $\delta v_L$, $\delta v_T$ as a function of separation. This is intended to convey the degree to which this flow differs from the non-relativistic, incompressible case. In particular, notice that the peak of the velocity distribution at $v \sim 0.14c$ corresponds to a Lorentz gamma factor $\gamma \sim 1.01$, while the largest velocity is $v \sim 0.52 c$, corresponding to $\gamma \sim 1.17$. The bulk of the flow can therefore be considered non-relativistic (for comparison, the sound speed for this $2+1$ dimensional conformal fluid is $c_s \sim 0.71c$, so the flow is also subsonic). The density distribution shows a standard deviation of $0.055$ and a peak at $0.97$. The velocity differences are roughly Gaussian distributed with zero mean, and their widths $\sigma$ are comparable to the sound speed. One may thus describe this flow as being in the \emph{weakly-compressible regime}. Notice that the density distribution peaks at a value less than $1$ and has a stronger tail at lower values, which means that a bias is formed in favour of under-density with respect to the initially uniform value of $1$. Given the characteristics of the flow described, employing the Newtonian energy and enstrophy 
to connect with known results in the Newtonian regime is justified.

The total specific Newtonian kinetic energy of a representative simulation is shown as a function of time in Fig.~\eqref{fig:Et-and-Wt}. The energy plateaus after approximately $t=20t_{LC}$, indicating a statistically steady-state. Correlation functions are computed at $t=25t_{LC}$. In order to obtain snapshots of the flow which are statistically independent, one can choose the temporal spacing between samples to be at least one large-eddy turnover time, determined through $T=U/L$, where $U$ is a typical large-scale speed and $L$ is the large length scale. We estimate $U$ by applying a low-pass Fourier filter to the velocity field at a representative time $t=25t_{LC}$, with all wavenumbers larger than the friction scale $k_\alpha$ being set to zero, then choosing $U$ as the mode of the resulting velocity distribution. The large length scale $L$ is chosen as $2\pi L_\alpha$. We find this procedure gives roughly $T=25t_{LC}$, which is the same amount of time required to evolve the fluid from rest to a steady-state. Thus, we opt to evolve the fluid from rest to obtain each flow realization, rather than evolve from a steady-state at time $t$ to a later time $t+T$. This reduces the risk that each flow realization is not statistically independent.

The spectral energy $E(k)$ and flux $\Pi(k)$ (averaged over $200$ flow realizations) are shown in Fig.~\eqref{fig:E(k)}. $\Pi(k)$ is computed using the formula $\Pi(k) = \left\langle \boldsymbol{v}_{k}^{<}\cdot (\boldsymbol{v}_k^{<} \cdot \boldsymbol{\nabla} \boldsymbol{v}_k^{>}) \right\rangle + \left\langle \boldsymbol{v}_{k}^{<}\cdot (\boldsymbol{v}_k^{>} \cdot \boldsymbol{\nabla} \boldsymbol{v}_k^{>}) \right\rangle$, as described in~\cite{Frisch:1996}. Here, the superscripts $>,<$ denote Fourier-filtered quantities with all wavenumbers set to zero below or above the given $k$, respectively. The familiar Kolmogorov scaling of the spectral energy $E(k) \propto k^{-5/3}$ seems to hold, and the spectral energy flux exhibits qualitative behaviour similar to that displayed in~\cite{Boffetta:2010}. The intersection of $\Pi(k)$ with the horizontal axis at $k=100$ indicates the injection of energy there, since energy is flowing away from that length scale. Negative values of $\Pi(k)$ for $k<100$ indicate the inverse-cascade of energy. 

\begin{figure}[h!]
\centering
\includegraphics[width=0.8\textwidth]{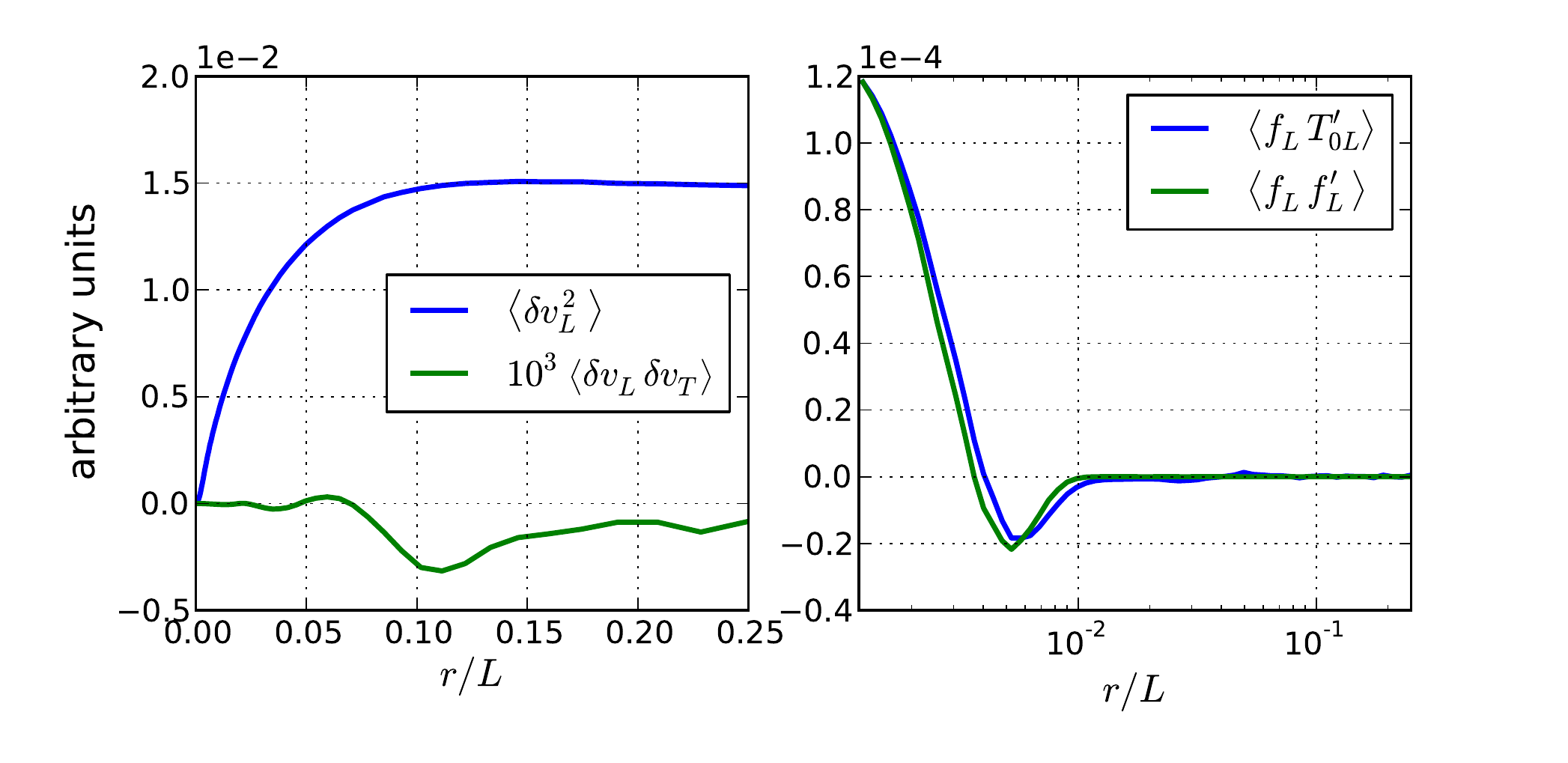}
\caption{Left: Numerical evidence of isotropy. The mixed correlation $\left\langle \delta v_L \delta v_T \right\rangle$ is supposed to vanish under isotropy, and it is measured to be less than the non-vanishing purely longitudinal correlation $\left\langle \delta v_L^2 \right\rangle$ by a factor of more than $10^3$. All correlations are computed over $10^4$ flow realizations. Right: Numerical evidence that $\left\langle f_i f^{\prime i} \right\rangle$ and $\left\langle f_i T_0^{\prime i} \right\rangle$ are proportional, in particular vanishing quickly with increasing $r$. Note that the former has been scaled by an overall constant in order to match with the latter at $r=0$. We show only the longitudinal correlations here - the transverse ones look the same.}\label{fig:isotropy-and-Novikov}
\end{figure}

To provide evidence that the flow is indeed statistically isotropic, we compute 2nd-order velocity correlation functions of purely longitudinal and mixed types, the latter of which is expected to vanish under isotropic conditions. Figure.~\eqref{fig:isotropy-and-Novikov} (left) illustrates the results obtained with an average over $10^4$ flow realizations, which shows the mixed type
being negligible with respect to the purely longitudinal case. 
 In addition, the right plot demonstrates that we have achieved $\left\langle f_i f^{\prime i} \right\rangle \propto \left\langle f_i T^{\prime i}_0 \right\rangle$, which is a crucial part of the derivation in Sec.~\eqref{sec:inversecascadescaling} of the linear scaling of $\left\langle T_{0i} T^{\prime ij} \right\rangle$ in the inverse-cascade range. However, the proportionality factor is on the order of $10^3$ rather than $2$ as the argument in Appendix~\eqref{app:Novikov} would suggest. This discrepancy is not surprising, as our force is not $\delta$-correlated in time as the argument in Appendix~\eqref{app:Novikov} requires. A proper numerical implementation of $\delta$-correlated statistics requires a modified integration algorithm, as described in~\cite{Honeycutt:1992}.

As a further display of the properties of this flow, we also compute velocity structure functions of orders $1$ through $4$, but with the absolute value of the velocity differences taken in the case of odd orders (this has been argued~\cite{Benzi:1993} to preserve the scaling properties, though it obscures the overall magnitude of the correlation). By taking the absolute value, all contributions to the correlation add constructively, which improves the convergence drastically (this is what was done in~\cite{Radice:2012pq} for a relativistic fluid in $d=3$). The scaling behaviour reflects the Kolmogorov expectation $\left\langle | \delta v_L |^n \right\rangle \propto r^{n/3}$ increasingly poorly as $n$ increases, but the same phenomenon has also been reported in~\cite{Boffetta:2010} for positive-definite velocity structure functions.

Lastly, it is interesting to closely examine the non-vanishing correlations, $\left\langle T^{\prime}_{0L} T_{LL} \right\rangle$ and $\left\langle T^\prime_{0T} T_{LT} \right\rangle$, together
with their incompressible limits, $(9/4)\left\langle v^\prime_L v_L v_L \right\rangle$ and $(9/4)\left\langle v^\prime_T v_L v_T \right\rangle$), respectively. This comparison will be discussed in the next section.

\begin{figure}[h!]
\includegraphics[width=0.9\textwidth]{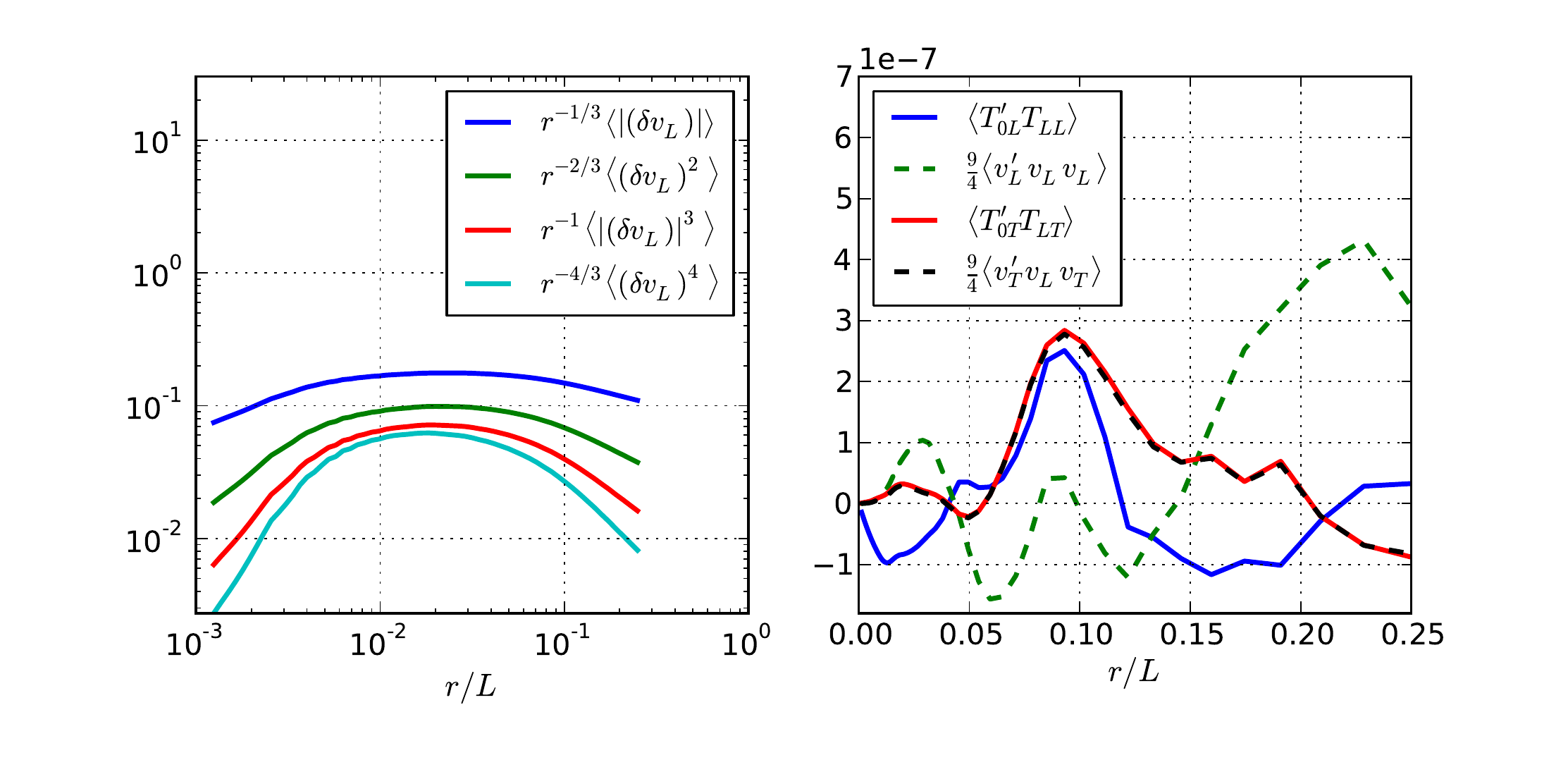}
\caption{Correlations plotted as functions of the separation $r/L$, with $L \equiv$ the size of the box. Left: Velocity structure functions of orders $n=1-4$, compensated by the Kolmogorov scaling $r^{n/3}$ (see~\cite{Boffetta:2012} for details about this expectation). The odd orders have an absolute value operation performed on $\delta v_L$, which causes them to converge more rapidly with sample size $N$. Right: The two relativistic correlations $\left\langle T^{\prime}_{0L} T_{LL} \right\rangle$ and $\left\langle T^{\prime}_{0T} T_{LT} \right\rangle$ which are not expected to vanish under isotropic conditions, along with their incompressible limits $(9/4)\left\langle v^{\prime}_L v_L v_L \right\rangle$ and $(9/4)\left\langle v^{\prime}_T v_L v_T \right\rangle$. The factors of $9/4$ are left over after the limit is taken. All correlations are computed over $10^4$ flow realizations with a grid size of $800^2$.}\label{fig:maincorrs} 
\end{figure}
%
%
\subsection{Discussion}
The correlations shown in Fig.~\eqref{fig:maincorrs} (right) are still unresolved, even with a sample size of $10^4$ flow realizations. We observe this by computing, at each $r/L$, the standard deviation of the sample divided by the square root of the number of samples, $\sigma/\sqrt{N}$. We find it to be comparable with the value of the correlation itself, and thus conclude that the fluctuations have not averaged down sufficiently. Poor signal-to-noise constitutes the main difficulty in numerically measuring odd-order correlations in compressible turbulence. For our current simulation, a much larger sample size is required, as we will discuss later.

\begin{figure}[h!]
\centering
\includegraphics[width=.7\textwidth]{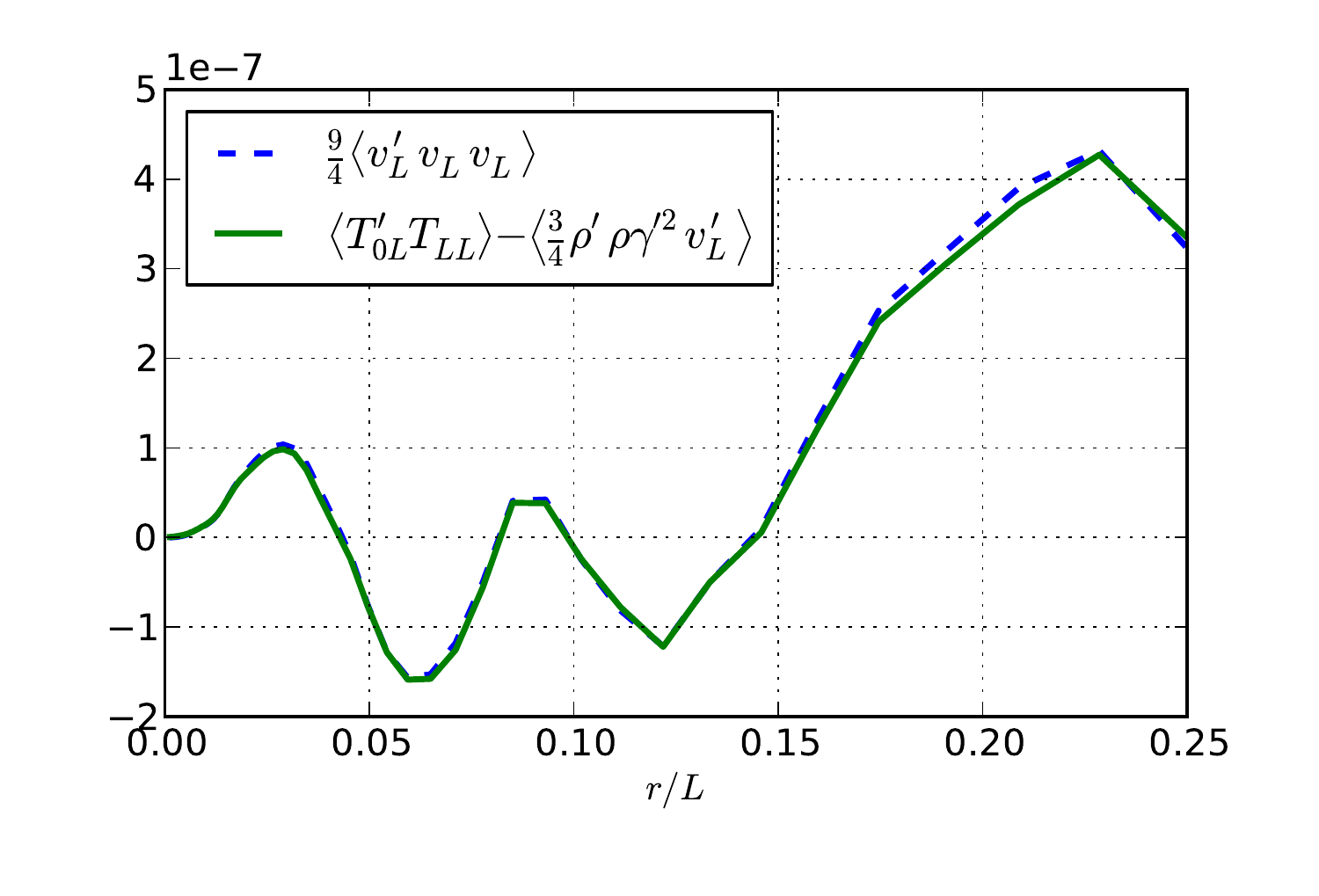}
\caption{The purely longitudinal relativistic correlation plotted without its spoiler term $\left\langle \frac{3}{4} \rho^\prime \rho {\gamma^\prime}^2 v^\prime_L \right\rangle$, as compared with its incompressible limit. The agreement is considerably improved, demonstrating that the spoiler term is sensitive to compressive effects.} \label{fig:LLLwout-spoiler}
\end{figure}

Nevertheless, some relevant conclusions can still be made. Firstly, notice in Fig.~\eqref{fig:maincorrs} (right) that there is a great disparity in how well $\left\langle T^{\prime}_{0T} T_{LT} \right\rangle$ and $\left\langle T^{\prime}_{0L} T_{LL} \right\rangle$ match with their incompressible limits. The former is indistinguishable from its limit in the plot, though zooming in reveals that there are small differences. The latter, on the other hand, bears little resemblance to its incompressible limit. To gain insight about this, consider the two correlations written in terms of the primitive variables:
\begin{eqnarray}
\left\langle T^{\prime}_{0L} T_{LL} \right\rangle &=& \left\langle \frac{9}{4} \rho^\prime \rho {\gamma^\prime}^2 \gamma^2 v^\prime_L v_L v_L \right\rangle + \left\langle \frac{3}{4} \rho^\prime \rho {\gamma^\prime}^2 v^\prime_L \right\rangle \label{eq:LLLexpansion}, \\
\left\langle T^{\prime}_{0T} T_{LT} \right\rangle &=& \left\langle \frac{9}{4} \rho^\prime \rho {\gamma^\prime}^2 \gamma^2 v^\prime_T v_L v_T \right\rangle \label{eq:TLTexpansion}.
\end{eqnarray}
In the incompressible limit, $\gamma,\rho \rightarrow 1$. The 2nd term on the right-hand side of Eq.~\eqref{eq:LLLexpansion} will therefore become $\propto \left\langle v^\prime_L \right\rangle$ which is zero by statistical symmetries. Indeed, we find numerically that the overall magnitude of $\left\langle \frac{3}{4} \rho^\prime \rho {\gamma^\prime}^2 v^\prime_L \right\rangle$ is roughly $10^5$ times larger than $\left\langle v_L^{\prime}\right\rangle$. This indicates that the underlying probability distribution for this term is highly sensitive to compressive effects, which cause it to become considerably wider, translating into much larger fluctuations. We can also implicate this term in the disagreement between Eq.~\eqref{eq:LLLexpansion} and its incompressible limit by subtracting it from Eq.~\eqref{eq:LLLexpansion}. We display the result in Fig.~\eqref{fig:LLLwout-spoiler}, where it is seen that the agreement improves considerably. It remains to be seen whether this spoiler term will average down to become negligible in the weakly compressible regime we are exploring here.

\begin{figure}[h!]
\centering
\includegraphics[width=\textwidth]{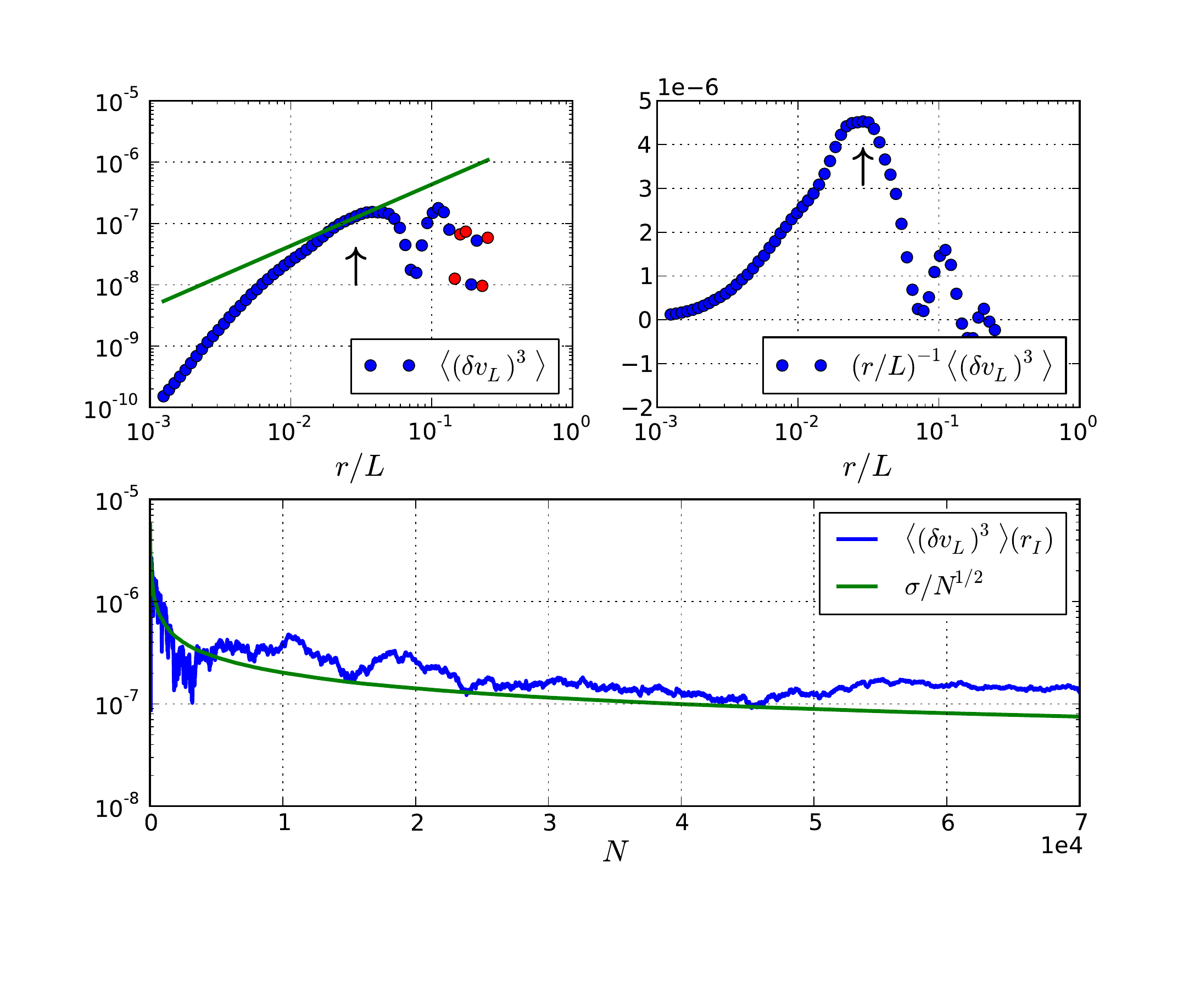}
\caption{Top row: $\left\langle (\delta v_L)^3 \right\rangle$ (left) and $(r/L)^{-1}\left\langle (\delta v_L)^3 \right\rangle$ (right) averaged over $7\times 10^4$ realizations of an incompressible flow at $800^2$ resolution. For visual comparison, a linear trend is displayed as a straight line on the left. The flat interval on the right plot also corresponds to a linear trend, which is a similar result to that of~\cite{Boffetta:2000}. Negative values have been indicated in red on the left. Bottom: the sample average and uncertainty $\sigma/N^{1/2}$ plotted versus sample size $N$ for the purely longitudinal velocity structure function $\left\langle (\delta v_L)^3 \right\rangle |_{r=r_I}$, for a separation $r_I$ in the inertial range (indicated with an arrow in the top row). This separation corresponds to a wavenumber of $k=35$ in grid units. The uncertainty reduces to roughly $1/2$ of the average at $N=7\times 10^4$, translating into a signal-to-noise ratio of $\sim 2$. } \label{fig:NSstrucs}
\end{figure}

Secondly, note that the third-order velocity correlation $\left\langle v^{\prime}_L v_L v_L \right\rangle$ has been successfully resolved in simulations of an exactly-incompressible fluid in~\cite{Boffetta:2000} when averaged over only tens of flow realizations, albeit with a larger grid size of $2048^2$. As a proof of principle, switching to the less-costly case of an exactly-incompressible fluid\footnote{Which incorporates a Poisson solver to impose incompressibility, and adopts a second-order ``white noise'' Runge-Kutta algorithm~\cite{Honeycutt:1992}.} we obtain similar results for our current grid size of  $800^2$, shown in Fig.~\eqref{fig:NSstrucs} after averaging over $7 \times 10^4$ flow realizations. An investigation into the dependence of the signal-to-noise of the correlations on compressive effects and the nature of the random external force is left for future studies. For such
work, it is important to estimate the sample size required to resolve a given correlation. Such an estimate
can be obtained in terms of the standard deviation of the underlying distribution and the scaling prediction. For instance, Eq.~\eqref{R1} provides the prediction for the strength of the signal. Supposing the underlying distribution for $T^\prime_{0i}T^i_L |_{r=r_I}$ has a standard deviation $\sigma(r_I)$, where $r_I$ is a separation within the inverse-cascade range. Then the signal-to-noise ratio $\text{SNR}$ would be given by
\begin{eqnarray}
\text{SNR} = \frac{\epsilon r_I}{2} \frac{N^{1/2}}{\sigma},
\end{eqnarray}
where $N$ is the sample size. Solving for $N$ yields
\begin{eqnarray}
N = \left(\frac{2\sigma}{\epsilon r_I}\text{SNR}\right)^2.
\end{eqnarray}

%
%
%
\section{Summary}\label{sec:summary}

In this work we derived 
scaling relations in fully-developed relativistic turbulence in two spatial dimensions.
We considered both the inverse- and direct-cascade ranges, and the relativistic results reduce in the non-relativistic limit
to the corresponding  scalings in the incompressible case. 
This derivation bridges
known results in the field of incompressible fluid turbulence with ongoing work in the relativistic case.

We have also begun a numerical experiment in an effort to measure the derived scaling relations through direct numerical simulations. We showed through Fig.~\eqref{fig:vrho-histograms} that the flow displays Mach numbers around $0.2$ in both absolute and relative velocities, and is thus weakly-compressible. In this regime, the probability distribution underlying the term $\left\langle \rho^\prime \rho \gamma^{\prime 2} v^\prime_L \right\rangle$ in Eq.~\eqref{eq:LLLexpansion} acquires a large standard deviation as compared with its incompressible counterpart, $\left\langle v_L^\prime \right\rangle$, being $10^5$ times larger with the same sample size. While the latter can be argued to vanish by isotropy, the former cannot. This opens the possibility that this `spoiler term' provides the dominant deviation of Eq.~\eqref{eq:LLLexpansion} from its incompressible limit to leading order in compressive effects, although this must be verified by increasing the sample size considerably. It would be interesting to observe deviations of the relativistic correlations derived here from their incompressible limits in the highly-compressible or relativistic regimes.

The signal-to-noise of odd-order correlations is the overarching difficulty in measuring them accurately. Indeed, previous studies of compressible Navier-Stokes turbulence (eg.~\cite{Porter:2002,Kritsuk:2007,Benzi:2008}) and relativistic turbulence (eg.~\cite{Radice:2012pq}) sidestepped this issue for the case of velocity structure functions by taking the absolute value of the velocity differences, $\left\langle |\delta v |^n \right\rangle$. The result of this procedure is that every term in the average adds constructively, and the scaling behavior has been argued in~\cite{Benzi:1993} to be preserved. In our case, an analogous work around is not available for the relativistic correlations in Eq.~\eqref{R1}, since the coupling of factors of the velocity in $T_{ij}$ prevent writing the correlations in terms of velocity differences. We are continuing our effort to resolve the correlations, and those results will be left for a future communication.

As a final comment, our work examining the behaviour of relativistic, conformal fluids undergoing turbulence has a natural connection with both conformal field theories and gravity through holography and the fluid-gravity correspondence (e.g.~\cite{Baier:2007ix,Bhattacharyya:2008jc,VanRaamsdonk:2008fp}).
The correspondence relates the fluid stress tensor in $d$ dimensions to the 
asymptotic behaviour of a $d+1$ dimensional black hole spacetime metric (up to counter-terms to obtain
a finite expession) as,
\begin{equation}
T_{ab} = \lim_{r \to\infty} \frac{r^d}{8\pi G_N^{(d+1)}} \left ( K_{ab} - K \gamma_{ab} \right ) \, ,
\end{equation}
where $\gamma_{ab}$ and $K_{ab}$ are the intrinsic and extrinsic curvatures respectively of a timelike 
surface at $r \to\infty$.
It implies that in the turbulent gravitational regime, correlations involving the metric tensor itself should obey scaling behaviour of the form discussed here\footnote{That would apply even in asymptotically flat scenarios~\cite{Yang:2014tla,Yang:2015jja}}. The implications of such an
intriguing observation 
are still unexplored.
%
%

\acknowledgements
We would like to thank G. Boffetta and D. Bernard for discussions, as well as
K. Cannon and S. Green for both discussions and comments throughout this project.
This work
was supported by NSERC through a Discovery Grant and by CIFAR (L.L.),
by the I-CORE program of Planning and Budgeting Committee (grant number 1937/12), the US-Israel Binational Science Foundation, GIF and the ISF Center of Excellence (Y.O.). J.R.W.S. acknowledges support from both NSERC CGS-M and OGS during the completion of this work.
Research at Perimeter
Institute is supported through Industry Canada and by the Province of
Ontario through the Ministry of Research \& Innovation.  
This research was enabled in part by support provided by Scinet and Compute Canada.

\appendix
\section{Relating $\left\langle T_{0i} f^{\prime i} \right\rangle$ and $\text{tr}F$} \label{app:Novikov}
Here we present an adaptation of an argument by Novikov~\cite{Novikov:1965}. In~\cite{Novikov:1965} it was shown that, for a homogeneous Gaussian random field $f_i(\boldsymbol{x},t)$ satisfying Eq.~\eqref{eq:beforeNovikov}, one can write its correlation with a functional $R[f]$ as
\begin{eqnarray}
\left\langle f_i(\boldsymbol{r},t) R[f] \right\rangle = \int F_{ik}(\boldsymbol{r}-\boldsymbol{r}^\prime) \left\langle \frac{\delta R[f]}{\delta f_k(\boldsymbol{r}^\prime,t)} \right\rangle d^3 r^\prime . \label{eq:Novikov_identity}
\end{eqnarray}
The strategy is to then regard $T_{0i}$ as a functional of the external force $f$, then compute its functional derivative and plug that into the above relation. To this end, one writes the equation of motion as
\begin{eqnarray}
T_{0i}(\boldsymbol{r},t) = T_{0i}(\boldsymbol{r},0) - \int_0^t \partial^j T_{ij} d\tau + \int_0^t f_i d\tau .
\end{eqnarray}
Upon applying the variational derivative at differing position and time $t^\prime$ such that $0 < t^\prime < t$, one obtains
\begin{eqnarray}
\frac{\delta T_{0i}(\boldsymbol{r},t)}{\delta f_k (\boldsymbol{r}^\prime,t^\prime)} = -\int_{t^\prime}^t \frac{\delta}{\delta f_k (\boldsymbol{r}^\prime,t^\prime)}\partial^j T_{ij} d\tau + \Theta (t-t^\prime) \delta_{i}^{k} \delta (\boldsymbol{r}-\boldsymbol{r}^\prime), \label{eq:functionalderivative}
\end{eqnarray}
where $\Theta$ is the step function with $\Theta(0) = 1/2$. The appearance of the step function and the change of the lower limit of integration from $0$ to $t^\prime$ is a physical requirement, namely that nothing can depend on the force evaluated at a future time. For instance, the integrand on the right-hand side evaluated at a time $\tau$ cannot depend on the force at a time $t^\prime > \tau$, and so the lower limit of integration cannot extend below $t^\prime$. Upon evaluating Eq.~\eqref{eq:functionalderivative} at equal time, the integral appearing on the right-hand side vanishes so long as the integrand remains finite, and one obtains
\begin{eqnarray}
\frac{\delta T_{0i}(\boldsymbol{r},t)}{\delta f_k (\boldsymbol{r}^\prime,t)} = \frac{1}{2} \delta_{i}^{k} \delta(\boldsymbol{r}-\boldsymbol{r}^\prime). \label{eq:functionalderivative2}
\end{eqnarray}
Setting $R[f] = T_{0i}(\boldsymbol{r}^{\prime\prime},t)$ and then using Eq.~\eqref{eq:functionalderivative2} and Eq.~\eqref{eq:Novikov_identity}, one obtains the desired result,
\begin{eqnarray}
\left\langle T_{0i} f^{\prime i} \right\rangle = \frac{1}{2}\text{tr}F.
\end{eqnarray}
%
%
\section{The forcing function}\label{sec:forcingfunction}
We wish to construct a Gaussian random forcing function which is divergence-free and statistically homogeneous and isotropic, whose two-point correlation decays quickly with increasing distance, and which is delta-correlated in time (sometimes called \emph{white noise in time}). In symbols, $\partial_i f^i =0$ and 
\begin{eqnarray}
\left\langle f_i (\boldsymbol{r},t^{\prime}) f_j (0,t) \right\rangle = F_{ij}(\boldsymbol{r}) \delta(t-t^{\prime}),
\end{eqnarray}
where $F_i^i \equiv \text{tr}F = \text{tr}F(r)$. 

This latter condition is the sense in which this is an isotropic vector field. For $f_i$ divergence-free, the stronger form of isotropy where $F_{xx} = F_{xx}(r)$ and $F_{yy}= F_{yy}(r)$ forces $F_{xx}=F_{yy}=0$. One can see this by moving to Fourier space, where the $xx$ and $yy$ two-point correlation functions become $\left\langle \hat{f}_x (\boldsymbol{k}) \hat{f}_x^{*} (\boldsymbol{k}) \right\rangle \equiv g(k)$ and $\left\langle \hat{f}_y (\boldsymbol{k}) \hat{f}_y^{*} (\boldsymbol{k}) \right\rangle \equiv h(k)$, for some functions $g,h$ of the magnitude of the wavevector only. The divergence-free condition reads $k_x\hat{f}_x + k_y \hat{f}_y = 0$, which allows us to convert between these correlation functions. Thus, for $k_x \neq 0$
\begin{eqnarray}
\left\langle \hat{f}_x \hat{f}_x^{*} \right\rangle = \left\langle \frac{k_y^2}{k_x^2}\hat{f}_y \hat{f}_y^{*} \right\rangle \nonumber
&=& \frac{k_y^2}{k_x^2} \left\langle \hat{f}_y \hat{f}_y^{*} \right\rangle,
\end{eqnarray}
which contradicts $\left\langle \hat{f}_y (\boldsymbol{k}) \hat{f}_y^{*} (\boldsymbol{k}) \right\rangle \equiv h(k)$ unless $h(k)=0$. This is why we chose to sum over $i$ in Sec.~\eqref{sec:inversecascadescaling}, since the divergence-free nature of the force played a role in the argument. 

In our simulations, to generate a divergence-free force we derive it from a stream function $\psi$ such that
$\boldsymbol{f}=(\partial_y \psi, -\partial_x \psi)$. We thus specify $\psi$ itself as a Gaussian random, homogeneous, isotropic scalar field which is delta-correlated in time. In symbols,
\begin{eqnarray}
\left\langle \psi(\boldsymbol{r},t^{\prime}) \psi (0,t) \right\rangle = \Psi(r) \delta(t-t^{\prime}), \label{eq:streamfncorr}
\end{eqnarray}
with $\Psi(r)$ a thin Gaussian function, which ensures a short correlation length. In practice, $\psi$ is built in Fourier space, where the reality condition $\hat{\psi}^{*}(\boldsymbol{k}) = \psi(-\boldsymbol{k})$ is imposed, and where each mode receives a complex amplitude drawn from zero-mean Gaussian distributions whose widths are given by $\hat{\Psi}^{1/2} (k)$. As constructed, $\psi$ satisfies Eq.~\eqref{eq:streamfncorr}. The force itself then has $\text{tr}F$ whose Fourier transform is a wide Gaussian weighted by $k^2$. In real space, $\text{tr}F$ behaves as is plotted in Fig.~\eqref{fig:isotropy-and-Novikov} (right). At each step in the Runge-Kutta integration this procedure is repeated anew, thus giving different individual realizations of the random force. This is the sense in which we have approximated a delta-correlation in time. A proper implementation requires a modified algorithm, as described in~\cite{Honeycutt:1992}.
%
%
\section{The consequences of isotropy}\label{app:consequencesisotropy}
In their seminal paper, Karman and Howarth \cite{Karman:1938} argue as follows (we reproduce their argument for $d=2$). Let the two points under consideration lie on the $x$-axis. We say that the $x$-direction is the \emph{longitudinal direction}, pointing directly between the two points, while we say that all other perpendicular directions are \emph{transverse directions}. The triple velocity correlation functions can be listed as
\begin{eqnarray}
&& \left\langle v_x(0) v_x(\vec{\boldsymbol{r}}) v_x(\vec{\boldsymbol{r}}) \right\rangle, \left\langle v_y(0) v_x(\vec{\boldsymbol{r}}) v_x(\vec{\boldsymbol{r}}) \right\rangle, \left\langle v_x(0) v_x(\vec{\boldsymbol{r}}) v_y(\vec{\boldsymbol{r}}) \right\rangle\nonumber\\
&& \left\langle v_y(0) v_x(\vec{\boldsymbol{r}}) v_y(\vec{\boldsymbol{r}}) \right\rangle, \left\langle v_x(0) v_y(\vec{\boldsymbol{r}}) v_y(\vec{\boldsymbol{r}}) \right\rangle, \left\langle v_y(0) v_y(\vec{\boldsymbol{r}}) v_y(\vec{\boldsymbol{r}}) \right\rangle\nonumber.
\end{eqnarray}
Now, both directions $\hat{\boldsymbol{y}}$ and $-\hat{\boldsymbol{y}}$ are transverse, and by isotropy (or parity-invariance in $d=2$) every correlation will be invariant under a switch between them. However, any correlation with an odd number of $y$-components will undergo a change of sign when the $y$-axis is inverted. Those correlations therefore vanish. The remaining correlations are
\begin{eqnarray}
\left\langle v_x(0) v_x(\vec{\boldsymbol{r}}) v_x(\vec{\boldsymbol{r}}) \right\rangle, \left\langle v_y(0) v_x(\vec{\boldsymbol{r}}) v_y(\vec{\boldsymbol{r}}) \right\rangle, \left\langle v_x(0) v_y(\vec{\boldsymbol{r}}) v_y(\vec{\boldsymbol{r}}) \right\rangle.
\end{eqnarray}
The incompressibility condition $\sum_i \partial_i v_i=0$ further implies that only one of these remaining three correlations is independent (see~\cite{Karman:1938}). In the relativistic case, there does not necessarily exist an analogous condition, so all three correlations might be independent.

The same arguments about sign-flipping apply in the case of homogeneous, isotropic turbulence in a special relativistic perfect fluid. For a correlation such as $\left\langle T_{0i} T_{ij}^{\prime} \right\rangle$, one can see for example with a perfect fluid energy-momentum tensor
\begin{eqnarray}
T_{ab} = (\rho + p)u_a u_b + p\eta_{ab}, \label{eq:perfectfluidT}
\end{eqnarray}
where $\vec{\boldsymbol{u}} = \gamma (1,\vec{\boldsymbol{v}})$, that for $i\neq j$ we will have $T_{ij} = (\rho+p)\gamma^2 v_i v_j$ undergo a change in sign when one of the $i$- or $j$-axes is inverted. On the other hand, if $i=j$, then $T_{ii} = (\rho + p)\gamma^2 v_i v_i + p\delta_{ii}$ does not change sign when the $i$-axis is inverted. Furthermore, $T_{0i}=(\rho+p)\gamma^2 v_i$ changes sign when the $i$-axis is inverted. Thus all the facts are in place to run the same arguments presented in~\cite{Karman:1938}. This allows us to conclude that the only non-vanishing correlations of this type are
\begin{eqnarray}
\left\langle T_{0L}T_{LL}^{\prime} \right\rangle , \left\langle T_{0T}T_{LT}^{\prime} \right\rangle , \left\langle T_{0L}T_{TT}^{\prime} \right\rangle ,
\end{eqnarray}
where $L$ and $T$ are the longitudinal and transverse directions, respectively.

\section{Vorticity Behavior: 2+1 case}\label{sec:vorticityappendix}

\subsection{Unforced case}
The goal here is to derive a relativistic equation for the vorticity as defined in~\cite{Carrasco:2012nf}.
In that reference, vorticity is defined as 
\begin{equation}
 \Omega_{\mu \nu} = \nabla_{[\mu} \rho^{1/d} u_{\nu]} \, ,\label{carter}
\end{equation}
and was shown to give rise to the conserved current,
\begin{equation}
J^{\mu} \equiv \rho^{-2/3} (\Omega^{\alpha \beta} \Omega_{\alpha \beta}) u^{\mu} \, ,
\end{equation}
for a conformal perfect fluid. To derive the a scaling relation for this case by following a strategy similar to Sec.~\eqref{sec:directcascadescaling}, we first
require an equation in conservation form for a quantity related to vorticity. 
 
Obviously the quantity $W^a \equiv \epsilon^{abc} \Omega_{bc}$
satisfies $\partial_a W^a = 0$. This fact can also be obtained from the following
argument. As discussed in \cite{Carrasco:2012nf}, in 2+1 dimensions $\Omega_{ab} = \epsilon_{abc} u^c \Omega$
where $ \Omega^2 \equiv \Omega_{ab}\Omega^{ab} $. It follows that
\begin{eqnarray}
\partial_a W^a \equiv \partial_a (\Omega u^a) &=& \Omega \partial_a u^a + u^a \partial_a \Omega \, ,\nonumber \\
 &=&  \Omega \partial_a u^a - \frac{\Omega}{2} \left ( \partial_a u^a - \frac{2}{3\rho} u^a \partial_a \rho \right ) \, ,\nonumber \\
 &=&  \Omega \partial_a u^a - \frac{\Omega}{2} \left ( \partial_a u^a 2 \right ) \, ,\nonumber \\
 &=& 0\, ;
\end{eqnarray}
where we have used in the first line the relation derived in  Appendix C of \cite{Carrasco:2012nf} to show $\partial_a J^a = 0$ and
in the last line the hydrodynamic conservation equation along the flow velocity,
\begin{equation}
u^{\mu} \partial_{\mu} \rho  =  -\frac{d}{d-1}\rho (\nabla_{\mu} u^{\mu})
\equiv -\frac{d}{d-1}\rho \Theta \, , \label{parallel}
\end{equation}

We therefore have,
\begin{equation}
\nabla_0 (\Omega u^0) + \nabla_i (\Omega u^i) = 0 \, , \label{vorticityconservation}
\end{equation}
with $\Omega = \left( \partial_{[a} (T u_{b]}) \partial^{[a} (T u^{b]}) \right )^{1/2} $ which reduces to
the standard vorticity equation in the Newtonian limit.
%
%
\subsection{Forced case}\label{sec:relvortforce}
Suppose now there is a force acting in the problem, so that the conservation
equation reads $\nabla_a T^{ab} = f^b$. Projecting this equation along $u^a$ and orthogonal to it gives,
\begin{eqnarray}
u^a \partial_a \rho &=& - \frac{3}{2} \rho \partial_a u^a - \frac{1}{2} u_a f^a \, , \\
u^a \partial_a u_b &=& - \frac{1}{3 \rho} P^a_b \partial_a \rho + \frac{1}{3\rho} P_{ba} f^a\, .
\end{eqnarray}

Now, using the above relations, we have from Eq.~\eqref{carter},
\begin{eqnarray}
u^a \Omega_{ab} &=& u^a \partial_{[a} (\rho^{1/3} u_{b]} ) \, , \nonumber \\
 &=& u^a \partial_a (\rho^{1/3} u_b) - u^a \partial_b (\rho^{1/3} u_a)\, , \nonumber \\
 &=& u_b u^a \partial_a \rho^{1/3} + \rho^{1/3} u^a \partial_a u_b + \partial_b \rho^{1/3}\, , \nonumber \\
 &=& P^a_b \partial_a \rho^{1/3} + \rho^{1/3} \left( -\frac{1}{3\rho} P^a_b \partial_a \rho + \frac{1}{3\rho} P_{ab} f^a \right) \, , \nonumber \\
 &=& \frac{\rho^{-2/3}}{3} P_{ab} f^a \, .
\end{eqnarray}

We are interested now in exploring the condition $\partial_a W^a = 0$ as in the previous section under the influence of a force.
To proceed, let us observe that
\begin{eqnarray}
\epsilon^{abc} \Omega_{bc} &=& \epsilon^{adc} \delta_d^b \Omega_{bc} \, ,\nonumber \\
&=&\epsilon^{adc} (P_d^b -u_d u^b) \Omega_{bc} \, ,\nonumber \\
&=&\epsilon^{adc} P_d^b \Omega_{bc} - \frac{\rho^{-2/3}}{3} \epsilon^{adc} u_d P_c^b f_b \, ,\nonumber \\
&=&\epsilon^{adc} P_d^b \Omega_{bc} + \frac{\rho^{-2/3}}{3} \epsilon^{adc} u_c P_d^b f_b\, .
\end{eqnarray}
We thus arrive at the equation,
\begin{equation}
0 = \partial_a \left( \epsilon^{adc} P_d^b \left( \Omega_{bc} + \frac{\rho^{-2/3}}{3} u_c f_b \right) \right)
\end{equation}
This equation however does not relate the vector $W^a$ in the way we sought, i.e. an equation of
the form $\partial_a W^a = F$. Nevertheless,
it does motivate what the right vorticity-related vector should be, 
namely ${\cal W}^a \equiv \epsilon^{adc} P_d^b \Omega_{bc}$ which,
from the derivation above, satisfies
\begin{eqnarray}
\partial_a {\cal W}^a &=& -\epsilon^{adc} \partial_a \left ( \frac{\rho^{-2/3}}{3} P_c^b  u_d f_b \right ) \, , \nonumber \\
                      &=& -\epsilon^{adc} \partial_a \left (  \frac{1}{3 T^2} P_c^b u_d f_b \right ) \, , \label{eq:forcedrelvorteqn}
\end{eqnarray}
where we have used $\rho=T^3$ in the second line. Notice that
\begin{eqnarray}
 {\cal W}^a  {\cal W}_a = \Omega_{ab} \Omega^{ab} + {\cal O}(f) \, ,
\end{eqnarray}
with ${\cal O}(f)$ denoting terms depending linearly or quadratically on $f^b$, thus in the
absence of forcing ${\cal W}^a  {\cal W}_a = \Omega_{ab} \Omega^{ab}$ and we recover the conservation
of vorticity, Eq.~\eqref{vorticityconservation}, as expected from this quantity.
%
%
\subsection{Scaling Argument}
A scaling argument in the direct-cascade range involving the relativistic vorticity ${\cal W}^a$
can also be made, following Sec~\eqref{sec:directcascadescaling} closely.
First, define the right-hand side of Eq.~\eqref{eq:forcedrelvorteqn} as a forcing term $\tilde{\mathcal{F}}= \epsilon^{acd}\partial_a \left( \frac{1}{3T^2} P^b_c u_d f_c \right)$, where we use a tilde to distinguish this force from the one in Sec~\eqref{sec:directcascadescaling}. This gives the equation of motion succinctly as
\begin{eqnarray}
\partial^a \mathcal{W}_a = \tilde{\mathcal{F}}.
\end{eqnarray}
Second, consider the steady-state condition $\partial^0 \left\langle \mathcal{W}_0 \mathcal{W}^{\prime}_0 \right\rangle = 0$, and apply the time derivative and use the equation of motion to obtain,
\begin{eqnarray}
\partial^i \left\langle \mathcal{W}_i \mathcal{W}^{\prime}_0 \right\rangle = -\left\langle \tilde{\mathcal{F}} \mathcal{W}^{\prime}_0 \right\rangle,
\end{eqnarray}
where $\partial^i$ stands for the derivative with respect to the separation $\boldsymbol{r}=\boldsymbol{r}_2-\boldsymbol{r}_1$. Lastly, notice that far below the forcing scale, $r \ll L_f$, the right-hand side is constant, $\left\langle \tilde{\mathcal{F}} \mathcal{W}^{\prime}_0 \right\rangle \approx \left\langle \tilde{\mathcal{F}} \mathcal{W}_0 \right\rangle \equiv \tilde{\mathcal{\epsilon}}$, so upon integration (using isotropy)
\begin{eqnarray}
\left\langle \mathcal{W}_i \mathcal{W}^{\prime}_0 \right\rangle = -\frac{\tilde{\mathcal{\epsilon}}}{2} r_i,
\label{R4}
\end{eqnarray}
which is valid in the direct-cascade range. Notice that it is more difficult to integrate this expression twice than it is for the expression Eq.~\eqref{eq:relvortcorr} due to the presence of the projector in the definition of $\mathcal{W}_a$, which prevents taking the derivative operator outside without picking up additional terms. This means that obtaining an $r^3$ scaling relation from this linear one is not as straightforward as in the case of Eq.~\eqref{eq:relvortcorr}. 
%
%
\section{Energy condensate} \label{sec:energycondensate}
\begin{figure}
\includegraphics[width=\textwidth]{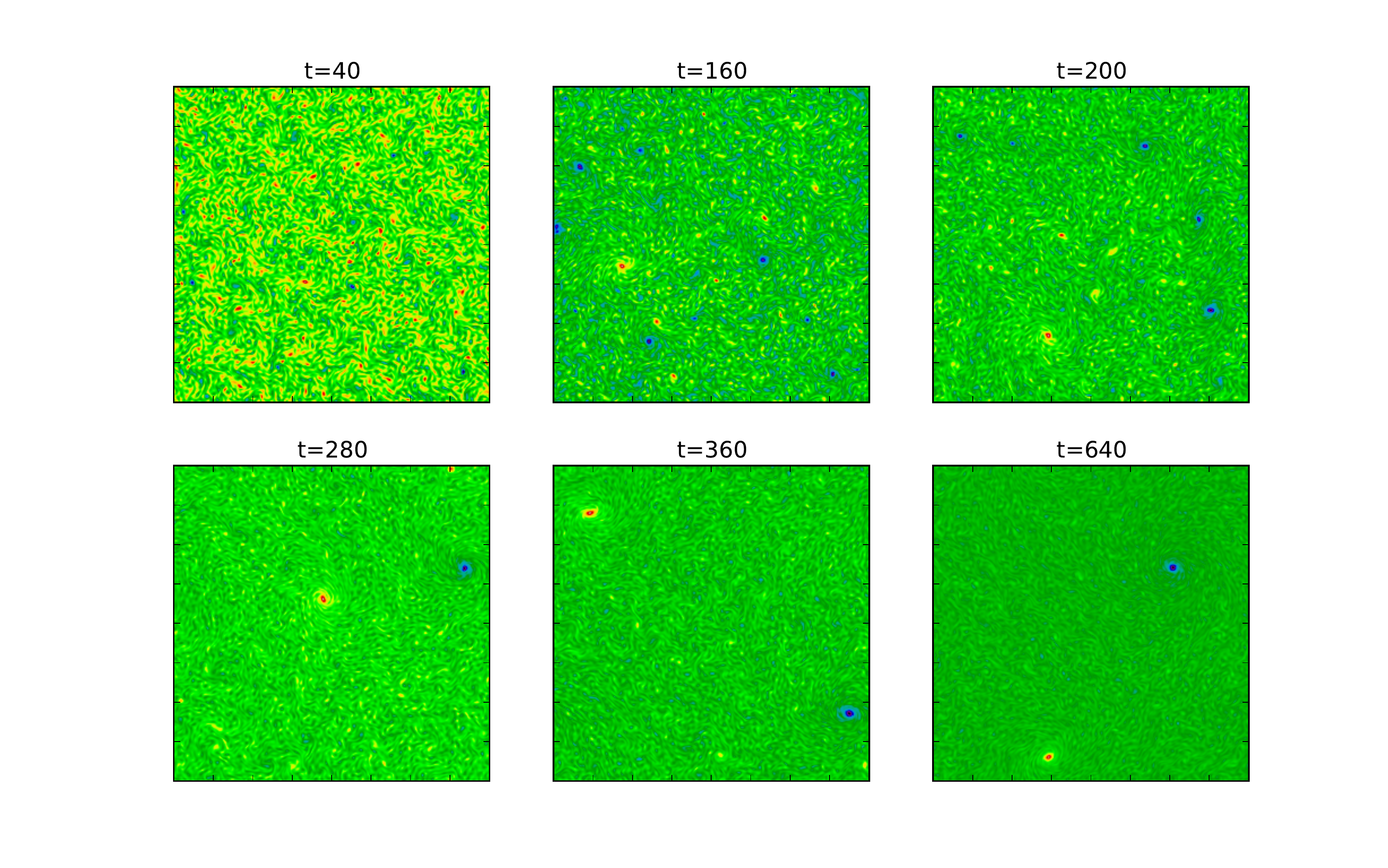}
\caption{Progression of the vorticity during the emergence of an energy condensate. The colour scale is omitted since this figure is only meant to qualitatively illustrate the anisotropic features of the energy condensate. All times are quoted in multiples of the light-crossing time $t_{LC}$.} \label{fig:vortprogression}
\end{figure}

In the absence of large-scale removal of energy in 2D, energy will build up in the gravest mode. Such a state is called an \emph{energy condensate}. For completeness, we present the
energy condensate and a method for removing it from
the analysis. As a concrete example, we adopt a periodic doman with grid size $N^2=400^2$ and a homogeneous, isotropic, random external force acting at $k_f=50$, normalized to a real-space amplitude $\beta=0.6$. After a sufficiently long time, the inverse
cascade leads to an energy condensate, as shown in Fig.~\eqref{fig:vortprogression}.
The figure displays the progression of the vorticity, with all times quoted in multiples of the light-crossing time $t_{LC}$. The colour scale has been omitted. Notice the late-time appearance of two dominant vortices of opposing sign.

\begin{figure}
\centering
\includegraphics[width=\textwidth]{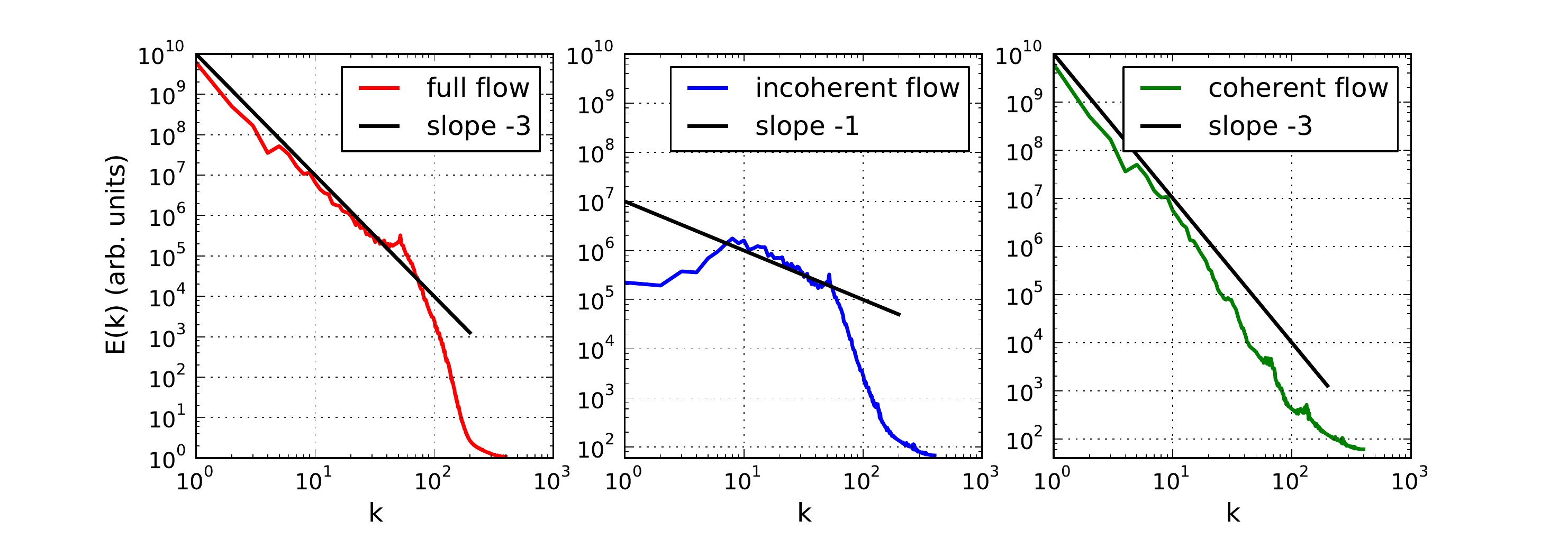}
\caption{Wavelet decomposition of the spectral energy of the condensate at $t=960 t_{LC}$ using Coiflet-12 wavelets and a threshold of $1$. Left: $E(k)$ for the full velocity field. Middle: $E(k)$ for the incoherent part of the velocity field. Right: $E(k)$ for the coherent part of the velocity field. The scaling behaviour of approximately $k^{-3}$, $k^{-1}$, and $k^{-3}$, respectively, is consistent with~\cite{Chertkov:2007}.} \label{fig:Ekcondenate}
\end{figure}

To analyze the resulting energy condensate one can make use of wavelets~\cite{Chertkov:2007}. 
We perform a similar analysis here as far as decomposing the velocity field into coherent and incoherent parts and computing the spectral energy of each. Fig.~\eqref{fig:Ekcondenate} illustrates the obtained results. The decomposition is performed using Coiflet-12 wavelets, which are a complete set of functions which are localized in both real and Fourier space. Their first two moments vanish (as well as their third and fourth moments), thus they couple weakly to Gaussian features. In other words, a relatively large number of basis elements with relatively low weights are required to represent Gaussian features of the data, whereas non-Gaussian features are represented by fewer basis elements with higher weights. Assuming the incoherent part of the velocity field is closer to Gaussian than the coherent part, one can therefore extract the incoherent part by imposing a threshold on the field in wavelet space, setting to zero all wavelet weights above a certain value, and then transforming back to real space. The remainder is the coherent part. 

\begin{figure}
\includegraphics[width=\textwidth]{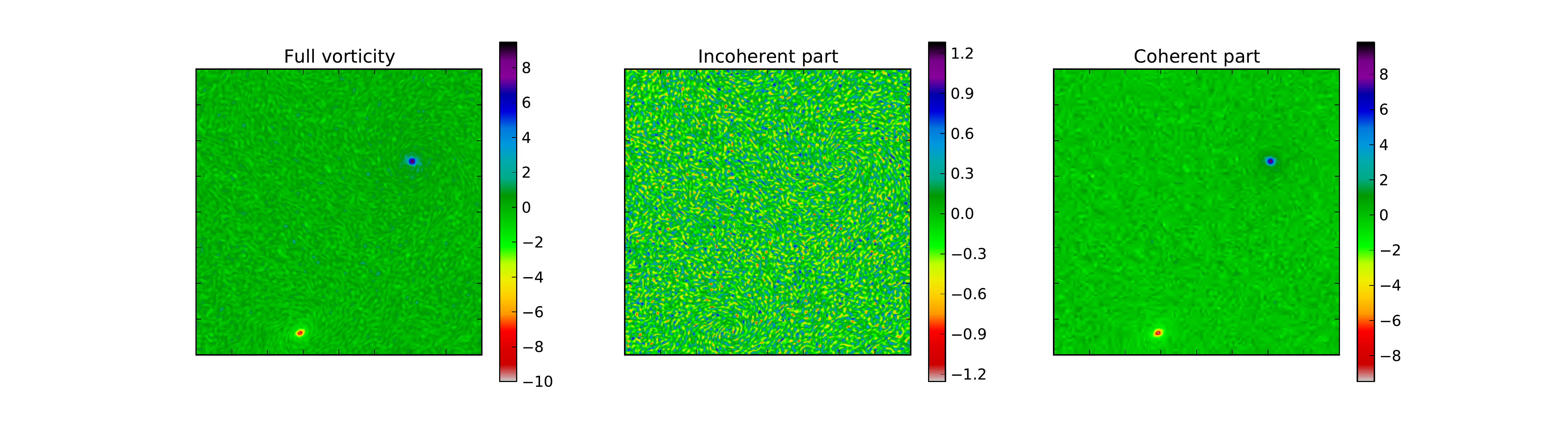}
\caption{Wavelet decomposition of the vorticity field at $t=960 t_{LC}$ with a threshold value of $3$. Left: the full vorticity field. Middle: the incoherent part. Right: the coherent part.} \label{fig:vortdecomp}
\end{figure}

To get a sense of what this procedure does, Fig.~\eqref{fig:vortdecomp} displays such a decomposition of the vorticity at $t=960 t_{LC}$ using a threshold value of $3$. Notice the increased blurriness of the coherent part of the vorticity (a common feature of compressed images, being represented by a small number of basis elements), and the dominant overall amplitude of the coherent  part with respect to the incoherent part. The energy scalings displayed in Fig.~\eqref{fig:Ekcondenate} are approximately consistent with~\cite{Chertkov:2007}. Note that there is a significant amount of arbitrariness in the choice of threshold value and wavelet type which we do not attempt to address here.

\bibliography{fluidbib}

\end{document}